\documentclass[aps,nofootinbib,superscriptaddress, showpacs,preprintnumbers,nofootinbibt,twocolumn]{revtex4-2}
\usepackage{epsfig}
\usepackage{multirow}
\usepackage{subcaption}
\usepackage{eurosym}
\usepackage{dcolumn}
\usepackage{bm}
\usepackage{enumerate}
\usepackage{float}
\usepackage{epstopdf}
\usepackage{amsmath}
\usepackage{bm}
\usepackage{amsfonts}
\usepackage{amssymb}
\usepackage{graphicx}
\usepackage{alphalph,mathtools}
\usepackage{etoolbox}
\usepackage{color}
\usepackage{booktabs}
\usepackage{hyperref}
\hypersetup{colorlinks,citecolor=blue}
\usepackage{footnote}
\usepackage{makecell,tabularx}

\def\be{\begin{equation}}
\def\ee{\end{equation}}
\def\bea{\begin{eqnarray}}
\def\eea{\end{eqnarray}}

\begin{document}

\title{\bf Constraining Cosmological Parameters with Viscous Modified Chaplygin Gas and Generalized Cosmic Chaplygin Gas Models in Horava-Lifshitz Gravity: Utilizing Late-time Datasets}
\author{Sayani Maity}
\email{sayani.maity88@gmail.com} 
\affiliation{Department of Mathematics, Sister Nivedita University, DG-1/2, Action Area 1, New Town, Kolkata-700 156, India.}
\author{Himanshu Chaudhary}
\email{himanshuch1729@gmail.com}
\affiliation{Pacif Institute of Cosmology and Selfology (PICS), Sagara, Sambalpur 768224, Odisha, India.}
\affiliation{Department of Applied Mathematics, Delhi Technological University, Delhi-110042, India} 
\affiliation{Department of Mathematics, Shyamlal College, University of Delhi, Delhi-110032, India.}
\author{Ujjal Debnath}
\email{ujjaldebnath@gmail.com} 
\affiliation{Department of
Mathematics, Indian Institute of Engineering Science and Technology, Shibpur, Howrah-711 103, India.}
\author{S. K. Maurya}
\email{sunil@unizwa.edu.om} 
\affiliation{Department of Mathematics and Physical Sciences, College of Arts and Sciences, University of Nizwa, Sultanate of Oman.}
\author{G.Mustafa}
\email{gmustafa3828@gmail.com} 
\affiliation{Department of Physics,
Zhejiang Normal University, Jinhua 321004, Peoples Republic of China}
\affiliation{New Uzbekistan University, Mustaqillik ave. 54, 100007 Tashkent, Uzbekistan}
\begin{abstract}
This study investigates accelerated cosmic expansion using the Viscous Modified Chaplygin Gas (VMMG) and Generalized Cosmic Chaplygin Gas (GCCM) within Horava-Lifshitz gravity. Our primary objective is to constrain essential cosmological parameters, such as the Hubble Parameter ($H_{0}$) and Sound Horizon ($r_{d}$). We incorporate recent datasets comprising 17 Baryon Acoustic Oscillation observations, 33 Cosmic Chronometer measurements, 40 Type Ia Supernovae data points, 24 quasar Hubble diagram data points, and 162 Gamma Ray Bursts data points. Additionally, we integrate the most recent determination of the Hubble constant (R22). We treat $r_{d}$ as a free parameter, which offers several advantages, including mitigating bias, enhancing precision, and improving compatibility with various datasets. Consequently, by introducing random correlations in the covariance matrix during simulation, errors are effectively reduced. Our estimated values of the Hubble constant ($H_0$) and $r_{d}$ consistently align with measurements from both the Planck and SDSS experiments. Additionally, cosmographic tests offer valuable insights into the dynamics of various cosmological models, enriching our understanding of cosmic evolution. Statefinder diagnostics provide deeper insights into cosmic expansion dynamics, aiding in distinguishing between both cosmological frameworks. Furthermore, the $o_{m}$ diagnostic test reveals that at late times, the VMMG model falls into the phantom region, while the Generalized GCCM falls into the quintessence region. Finally, the Akaike Information Criterion (AIC) and Bayesian Information Criterion (BIC) provide support for all models under consideration, indicating that each model offers a plausible explanation. Notably, the $\Lambda$CDM model emerges with the lowest AIC score, suggesting its relatively superior fit compared to others. Additionally, validation through the reduced $\chi_{\text{red}}^{2}$ statistic confirms satisfactory fits across all models, further reinforcing their credibility in explaining the observed data.
\end{abstract}
\maketitle
\tableofcontents
\section{Introduction}
In modern cosmology, explaining cosmic accelerated expansion presents a significant challenge. Observations from Type Ia supernovae (SNIa), baryon acoustic oscillations (BAOs), and the Cosmic Microwave Background Radiation (CMBR) reveal two distinct eras of accelerated expansion: inflation and late-time cosmic acceleration. This acceleration contradicts expectations of gravitational deceleration from normal baryonic matter. To account for this, physicists propose the existence of an unknown entity, termed dark energy (DE), responsible for producing a repulsive force. DE dominates 70\% of the total cosmic budget. The simplest model for DE arises from Einstein's field equations, known as the cosmological constant. This constant accounts for the observed acceleration, providing a theoretical framework for understanding cosmic accelerated expansion. The model faces two significant challenges known as fine-tuning and the cosmic coincidence problem~\cite{1,2,3,4}. To address these issues, researchers have pursued different avenues. One approach involves modifying the matter component in Einstein's field equations, resulting in various dark energy models such as the rolling scalar field~\cite{5,6}, tachyon~\cite{7}, and phantom models~\cite{8}. These models often stem from theories beyond classical physics, including quantum gravity models like holographic dark energy (HDE)~\cite{9} and new agegraphic dark energy (NADE)~\cite{10,11,12,13,14}. Another strategy focuses on altering the geometry part of Einstein's equations or extending General Relativity with quantum corrections. This leads to diverse modified gravity models, including $f(R)$ gravity~\cite{29} and $f(T)$ gravity~\cite{41} etc.\\\\
One of the compelling alternatives to Einstein's General Theory of Relativity is the Ho\v{r}ava-Lifshitz (HL) gravity. HL gravity offers a compelling alternative to Einstein's General Theory of Relativity, initially proposed by Petr Ho\v{r}ava. It aims to bridge the gap between general relativity and quantum mechanics by employing techniques from traditional quantum field theories to quantize gravity. Unlike general relativity, HL gravity exhibits full Lorentz symmetry emergence only at the infrared limit. It is characterized by power counting, making it renormalizable and potentially providing a UV-complete theory of gravity. Notably, this theory demonstrates anisotropic scaling of space and time dimensions. Various cosmological scenarios have been explored within the framework of HL gravity in the scientific literature. The investigation of scalar and tensor perturbations within Horava gravity, both with and without detailed balance, has been conducted on a flat background, as discussed in~\cite{26}. Furthermore, HL cosmology has been reexamined by introducing an additional scalar field, leading to an effective phase of dark energy~\cite{16}. Another study~\cite{32} delves into the Power Law Entropy Corrected Holographic Dark Energy (PLECHDE) model, employing the Granda-Oliveros cut-off as an infrared cut-off within the context of HL Horava cosmology. Additionally, the New Agegraphic Dark Energy (NADE) model, incorporating power-law corrected entropy, has been explored within the framework of HL cosmology. In addressing late-time acceleration of the Universe, various cosmological quantities have been examined for Tsallis, Rényi, and Sharma–Mittal holographic dark energy models, alongside modified field equations for logarithmic and power law versions of entropy corrected models, all within the framework of HL gravity~\cite{14} Moreover, the phenomenon of matter–antimatter asymmetry has been scrutinized in the background of HL gravity via gravitational baryogenesis~\cite{34} In \cite{Pourhassan:2022nay} a non-perturbative quantum correction in the black hole entropy has been considered and studied the consequence of thermodynamics of the Ho\v{r}ava-Lifshitz black hole at quantum scales. While these theoretical models may offer alternatives to the standard $\Lambda$CDM model, their viability hinges on their alignment with observational data sets such as TORNY and the Gold sample data set~\cite{17,18}. Consequently, numerous studies have been conducted in the literature to constrain dark energy and modified gravity models in light of observational data~\cite{30,31,45,46}. In~\cite{27}, researchers have investigated constraints on the parameter $\Lambda$ in HL gravity, which governs the transition from Ultra-Violet to Infra-Red regimes. Additionally, they considered the requirements of Big Bang Nucleosynthesis (BBN) in their analysis. The study on cosmological constraints of Horava gravity was further extended in~\cite{20} to incorporate observations from GW170817 and GRB170817A, while also addressing degeneracy with massive neutrinos. Numerous investigations have focused on constraining Horava gravity, utilizing different approaches including BBN bounds, and observational data from sources like CMB~\cite{21,22}, BAO~\cite{21,22}, galaxy power spectrum, and SNIa measurements \cite{23,24,25}. Consequently, in~\cite{43}, a cosmographic analysis was conducted on two recent parametrizations for dark energy models, namely CBDRM-type and CADMM-type, within the framework of HL gravity. In recent times, the Chaplygin gas dark energy model has gained significant traction in cosmology for its potential as a unified explanation for both the late-time cosmic acceleration and the matter-dominated era. In its simplest form, this model is characterized by an Equation of State (EoS). 
\begin{equation}
 p_d=-\frac{A}{\rho_d}, ~~~A>0 \mbox{ constant}
\end{equation}
Kamenshchik et al.~\cite{42} and Gorini et al.~\cite{44} introduced models that explain the late-time acceleration of the Universe but fall short in describing the era of structure formation. In response, various generalizations have been proposed in the literature. These include the Modified Chaplygin Gas (MCG)~\cite{15}, Generalized Chaplygin Gas (GCG)~\cite{36}, Generalized Cosmic Chaplygin Gas (GCCG)~\cite{61}, Variable Modified Chaplygin Gas (VMCG)~\cite{37}, New Variable Modified Chaplygin Gas (NVMCG)~\cite{38} and other~\cite{47,48,49,50}. Exploiting the idea that the Chaplygin gas may has viscosity, GCG with bulk viscosity has been introduced in \cite{ZHAI_2006}. Then this model was extended to Viscous Modified Chaplygin Gas and Viscous Modified Cosmic Chaplygin Gas in \cite{Saadat2013}. These models aim to unify dark matter and dark energy components in the Universe. Among these, the GCG model has shown promise by successfully aligning with observational data from sources such as the Wilkinson Microwave Anisotropy Probe (WMAP), CMBR, and BAO~\cite{28}. Various cosmologically significant studies have been done in literature by taking the members of Chaplygin gas family. For example: in \cite{Debnath_2022} a model of charged AdS black hole has been constructed by taking modified cosmic Chaplygin gas (MCCG). With the Assumption of a negative cosmological constant as a thermodynamics pressure, asymptotically charged AdS black hole thermodynamics has been examined with MCCG and produced a new solution to Einstein's AdS black hole field equations. In \cite{Ali2012} the cosmological scenarios of generalized Chaplygin gas has been investigated by taking both detailed and non-detailed balance version of gravitational background in the context of Hořava-Lifshitz gravity. The observational data from Type Ia Supernovae, Baryon Acoustic Oscillations and Cosmic Microwave Background, along with requirements of Big Bang Nucleosynthesis have been explored to constrain the parameters of the constructed model. In~\cite{39}, the Modified Chaplygin Gas (MCG) model has been explored concerning the peak location of the CMBR spectrum. The analysis suggested that the optimal range for the first parameter lies between $(-0.35, 0.025)$. Moreover, in~\cite{40}, the MCG model has been constrained using recent observational data from 182 Gold Type Ia Supernovae, the 3-year WMAP CMB shift parameter, and the SDSS baryon acoustic peak. The parameter values resulting in the best fit are determined as (-0.085, 0.822, 1.724) for parameters $A, B, \alpha$ respectively. In~\cite{51} presents the best-fit parameter values of the MCG model within the framework of Einstein-Aether Gravity. This determination is made through an analysis of BAO and CMB observations. Recently, in ~\cite{HCUDSKJNUMGFSKM24}, some of our authors have investigated the parametrizations of the dark energy equation of state (EoS) in the context of Horava-Lifshitz gravity. Mainly they have focused on the comparison of these parametrizations with the observational data collected from recent measurements of the Hubble parameter, H(z), cosmic chronometers, Type Ia Supernovae, Gamma-Ray Bursts (GRB), Quasars and uncorrelated Baryon Acoustic Oscillations (BAO) etc, which yields various constrains on the cosmic parameters. In \cite{1HCNUMMKUDGM24}, two dark energy models have been constructed with Linear and CPL parametrizations of the equation of state parameter in background of Horava-Lifshitz Gravity. By empolying various data set collected from cosmic chronometers (CC),Gamma-ray Burst (GRB), Type Ia Supernovae (SNIa), Baryon Acoustic Oscillations (BAO), Quasar (Q) and Cosmic Microwave Background Radiation (CMB), different constraints on the model parameter have been generated and also the present values of the important cosmological parameters such as $H_0$,$\Omega_{m_0},\Omega_{k_0}$ and $\Omega_{\Lambda_0}$ have been determined. In~\cite{19}, the authors explore observational constraints on MCG within the framework of HL gravity. They employ various observational datasets such as $H(z)$ - $z$, BAO peak parameter, and CMB shift parameter to scrutinize cosmologies both within detailed balance and beyond detailed-balance scenarios. Inspired by these investigations, we delve into the constraints on fundamental cosmic parameters within dark energy models. Specifically, we utilize the Pacif parametrization schemes, as introduced in \cite{pacif1,pacif2,pacif3}, along with recent measurements of the Hubble parameter, denoted as $H(z)$, encompassing 33 data points. For our analysis, we focus on two dark energy models: the GCCG~\cite{61} and the VMG ~\cite{47,48,49,50} within the framework of HL Gravity.\\\\ 
Motivated by the aforementioned considerations, we investigate two dark energy models within the framework of HL Gravity: Viscous Modified Chaplygin Gas (VMMG) and Generalized Cosmic Chaplygin Gas (GCCG). We utilize the covariance matrix to minimize errors and obtain the best-fit values for each cosmological parameter, including the present-day Hubble function $H_{0}$ and the Sound Horizon $r_{d}$. This paper is structured as follows: In Section \ref{Sec1}, we lay the groundwork by discussing the fundamental equations governing HL Gravity.  In Section \ref{Sec2}, we introduce two dark energy models, namely, VMMG and GCCG. Subsequently, we derive the expression for the Hubble parameter $H$ as a function of observational parameters and redshift $z$ for the two dark energy models under consideration. This step is crucial as it establishes a vital link between theoretical predictions and observational data. In Section \ref{Sec3}, we outline our methodology for analyzing the data and determining the optimal values of key cosmological parameters within both dark energy models.  In Section \ref{Sec4}, we delve into the discussion and visualization of cosmographic parameters associated with each dark energy model. In Section \ref{Sec5}, we perform statefinder and diagnostic tests to further assess the viability and compatibility of our models with observational data.  Section \ref{Sec6} presents the results obtained from our analysis, including the constraints on cosmological parameters, comparisons between models, and implications for our understanding of dark energy and gravitational theories. Finally, in Section \ref{Sec7}, we conclude our study by summarizing the key findings and offering concluding remarks on the implications of our results.
%%%%%%%%%%%%%%%%%%%%%%%%%%%%%%%%%%%%%%%%%%%%%%%%%%%%%%%%%%%%%%%%%%%%%%%
\section{Fundamental Equations of Horava-Lifshitz Gravity}\label{Sec1}
The Arnowitt-Deser-Misner (ADM) decomposition of the metric, often employed for its convenience, is expressed as \cite{52,53,54}:
\begin{equation}\label{HL1}
ds^2 = -N^2 dt^2 + g_{ij} \left(dx^i + N^i dt\right)\left(dx^j + N^j dt\right)
\end{equation}
Here, \( N \) denotes the lapse function, \( N_i \) the shift vector, and \( g_{ij} \) the metric tensor. A scaling transformation of coordinates \( t \rightarrow l^3 t \) and \( x^i \rightarrow l x^i \) is applied.
The Holst–Lorentz (HL) gravity action is comprised of two components: the kinetic and potential terms, represented as:
\[
S_g = S_k + S_v = \int dt d^3 x \sqrt{g} N \left( L_k + L_v \right)
\]
The kinetic term \( S_k \) is defined as:
\[
S_k = \int dt d^3 x \sqrt{g} N \left[ \frac{2\left(K_{ij}K^{ij} - \lambda K^2\right)}{\kappa^2} \right]
\]
Where the extrinsic curvature \( K_{ij} \) is given by:

\[
K_{ij} = \frac{\dot{g}_{ij} - \nabla_i N_j - \nabla_j N_i}{2N}
\]
When dealing with the Lagrangian, denoted as \( L_v \), the number of invariants can be decreased owing to its symmetric nature, as noted in references \cite{55,56,57}. This symmetry, referred to as detailed balance, influences the expanded expression of the action.
\begin{widetext}
$$S_g= \int dt d^3x \sqrt{g} N \left[\frac{2\left(K_{ij}K^{ij}
-\lambda K^2\right)}{\kappa^2}+\frac{\kappa^2
C_{ij}C^{ij}}{2\omega^4} -\frac{\kappa^2 \mu \epsilon^{i j k }
R_{i, j} \Delta_j R^l_k}{2\omega^2
\sqrt{g}}\right.$$$$\left.+\frac{\kappa^2 \mu^2 R_{ij} R^{ij}}{8}
-\frac{\kappa^2
\mu^2}{8(3\lambda-1)}\left\{\frac{(1-4\lambda)R^2}{4} +\Lambda R
-3\Lambda^2 \right\}\right],$$ where $C^{ij}=\frac{\epsilon^{ijk}
\Delta_k\left(R_i^j-\frac{R}{4} \delta^j_i\right)}{\sqrt{g}}$
\end{widetext}
In HL theory, the Cotton tensor and its covariant derivatives are derived with respect to the spatial metric \( g_{ij} \). Here, \( \epsilon^{ijk} \), \( \lambda \) are an antisymmetric unit tensor and a dimensionless constant, respectively. Constants \( \kappa \), \( \omega \), and \( \mu \) are also involved. Horava proposed a gravitational action under the assumption that the lapse function \( N \) depends only on time (i.e., \( N \equiv N(t) \)). When employing the Friedmann-Robertson-Walker (FRW) metric with \( N = 1 \), \( g_{ij} = a^2(t)\gamma_{ij} \), and \( N^i = 0 \), where
\[ \gamma_{ij}dx^i dx^j = \frac{dr^2}{1-kr^2} + r^2 d\Omega_2^2, \]
representing open (\( k = -1 \)), closed (\( k = 1 \)), and flat (\( k = 0 \)) universes, respectively. The Friedmann equations, governing the variation of \( N \) and \( g_{ij} \), take the following form \cite{58,59}:
\begin{widetext}
\begin{equation}\label{HLFriedmann1}
H^2=\frac{\kappa^2\rho}{6\left(3\lambda-1\right)}
+\frac{\kappa^2}{6\left(3\lambda-1\right)}\left[\frac{3\kappa^2\mu^2
k^2} {8\left(3\lambda-1\right)a^4}+\frac{3\kappa^2\mu^2 \Lambda^2}
{8\left(3\lambda-1\right)}\right]-\frac{\kappa^4 \mu^2 \Lambda
k}{8\left(3\lambda-1\right)^2a^2},
\end{equation}
\begin{equation}\label{HLFriedmann2}
\dot{H}+\frac{3H^2}{2}=-\frac{\kappa^2
p}{4\left(3\lambda-1\right)} -\frac{\kappa^2}
{4\left(3\lambda-1\right)}\left[\frac{3\kappa^2\mu^2 k^2}
{8\left(3\lambda-1\right)a^4}+\frac{3\kappa^2\mu^2 \Lambda^2}
{8\left(3\lambda-1\right)}\right]-\frac{\kappa^4 \mu^2 \Lambda
k}{8\left(3\lambda-1\right)^2a^2}
\end{equation}
\end{widetext}
In this context, the constant term is attributed to the cosmological constant, which is a distinctive contribution of HL (Horava-Lifshitz) gravity. Additionally, there's a term proportional to $\frac{1}{a^4}$, interpreted as the "Dark radiation term" according to~\cite{53,54}. Here, $H = \frac{\dot{a}}{a}$ represents the Hubble parameter, with the dot indicating a derivative with respect to cosmic time $t$. Considering a universe composed of two primary components: dark matter (DM) and dark energy (DE), the total energy density $\rho$ and total pressure $p$ are expressed as the sum of densities and pressures of these constituents: $\rho = \rho_m + \rho_d$ and $p = p_m + p_d$, respectively. Consequently, the conservation equations governing the evolution of DM and DE are as follows:
\begin{equation}\label{DM}
    \dot{\rho}_m + 3H(\rho_m + p_m) = 0
\end{equation}
and
\begin{equation}\label{DE}
    \dot{\rho}_d + 3H(\rho_d + p_d) = 0.
\end{equation}
In the context where dark matter is considered pressureless ($p_m = 0$), Equation (DM) implies that the density of dark matter, $\rho_m$, follows a scaling behavior of $\rho_m = \rho_{m0}a^{-3}$, where $\rho_{m0}$ represents the present energy density of dark matter, and $a$ denotes the scale factor. Now, let's introduce the EoS parameter, denoted by $w(z)$, defined as the ratio of pressure $p$ to energy density $\rho$. Utilizing Equation (DE), we can express the energy density of dark energy, $\rho_d$, as $\rho_{d0}~e^{3\int \frac{1+w(z)}{1+z} dz}$, where $\rho_{d0}$ signifies the present value of the energy density of dark energy. To further simplify, we introduce $G_{c}$ as $G_{c}=\frac{\kappa^2}{16\pi \left(3\lambda-1\right)}$, while ensuring that $\frac{\kappa^4 \mu^2 \Lambda}{8\left(3\lambda-1\right)}=1$ to maintain detailed balance. With these definitions, the Friedmann equations can be re-expressed accordingly.
\begin{equation}\label{H1}
H^2=\frac{8\pi G_c}{3}\left(\rho_m +
\rho_{d}\right)+\left(\frac{k^2} {2\Lambda
a^4}+\frac{\Lambda}{2}\right)-\frac{k}{a^2},
\end{equation}
\begin{equation}\label{H2}
\dot{H}+\frac{3}{2}H^2=-4\pi G_c p_d -\left(\frac{k^2}{4\Lambda
a^4}+\frac{3\Lambda}{4}\right)-\frac{k}{2a^2}.
\end{equation}
Using the dimensionless parameters $\Omega_{i0}\equiv\frac{8\pi
G_c}{3H_0^2}\rho_{i0}$, $\Omega_{k0}=-\frac{k}{H_0^2}$,
$\Omega_{\Lambda 0}=\frac{\Lambda}{2H_0^2}$, we obtain
\begin{widetext}
\begin{equation}\label{E}
H^2(z)=H_0^2\left[\Omega_{m0}(1+z)^3+\Omega_{k0}(1+z)^2
+\Omega_{\Lambda 0}+\frac{\Omega_{k0}^2(1+z)^4}{4\Omega_{\Lambda
0}}+\Omega_{d0}~e^{3\int \frac{1+w(z)}{1+z} dz}\right]
\end{equation}
\end{widetext}
with
\begin{equation}\label{Om}
\Omega_{m0}+\Omega_{d0}+\Omega_{k0}+\Omega_{\Lambda
0}+\frac{\Omega_{k0}^2}{4\Omega_{\Lambda 0}}=1
\end{equation}
The observational data analysis for linear, CPL and JBP models in
HL gravity have been studied in \cite{60}.

\section{Two Dark Energy Models}\label{Sec2}
\subsection{Viscous modified Chaplygin gas (VMMG)}
In the context of the viscous modified Chaplygin gas (VMCG), the equation describing pressure is provided as follows:
\begin{equation} \label{pressure_VMCG}
    p_d= A\rho_d-\frac{B}{\rho_d^\alpha}- 3\zeta_0 \sqrt{\rho_d}H
\end{equation}
In the given scenario, where $A$, $B$, and $\alpha$ are constants, substitution of $p_d$ from Eq. (\ref{pressure_VMCG}) into Eq. (\ref{DE}) yields:
\begin{equation}
    \rho_d= \left(\frac{B}{1+A-\sqrt{3}\zeta_0}+ \frac{C_1}{a^{3(1+\alpha)(1+A-\sqrt{3}\zeta_0)}} \right)^{\frac{1}{1+\alpha}}
\end{equation}
where $C_1$ is an integrating constant. The above expression can
be further re-written as:
\begin{equation}\label{rh}
    \rho_d= \rho_{d0}\{A_s+ (1-A_s)(1+z)^{3(1+\alpha)(1+A-\sqrt{3}\zeta_0)} \}^{\frac{1}{1+\alpha}}
\end{equation}
where $\rho_0$ being the energy density value at the present
epoch, $A_s= \frac{B}{(1+A-\sqrt{3}\zeta_0)C_1+B}$ satisfying the
conditions $0<A_s<1$ and $1+A-\sqrt{3}\zeta_0 >0$, and
$\rho_{d0}^{1+\alpha}= \frac{(1+A-\sqrt{3}\zeta_0)C_1+B}{1+A-\sqrt{3}\zeta_0}$, So from (\ref{E}), (\ref{Om}) and (\ref{rh}), we obtain
\begin{widetext}
\begin{eqnarray}
H^2(z)= && H_0^2\left[\Omega_{m0}(1+z)^3+\Omega_{k0}(1+z)^2
+\Omega_{\Lambda 0}+\frac{\Omega_{k0}^2(1+z)^4}{4\Omega_{\Lambda
0}} \right.  \nonumber
\\
&&+\left. \left(1-\Omega_{m0}-\Omega_{k0}-\Omega_{\Lambda
0}-\frac{\Omega_{k0}^2}{4\Omega_{\Lambda 0}}\right)\left[A_s+
(1-A_s)(1+z)^{3(1+\alpha)(1+A-\sqrt{3}\zeta_0)}
\right]^{\frac{1}{1+\alpha}}\right]
\nonumber\\
\end{eqnarray}
\end{widetext}

\subsection{Generalized Cosmic Chaplygin Gas (GCCG)}
In this analysis, we examine the generalized cosmic Chaplygin gas (GCCG) as a potential candidate for dark energy (DE), characterized by its EoS given by
\cite{61}
\begin{equation}\label{GCCG}
p_{d}=-\rho_{d}^{-\mu}\left[ K +
(\rho_{d}^{1+\mu}-K)^{-\eta}\right]
\end{equation}
where  $K=\frac{\nu}{1+\eta}-1$ and $-l<\eta<0$ with $l~(>1)$ is a constant. $\mu,~\nu$ are also constants. It can be seen that in limiting case
$\eta\rightarrow 0$, the GCCG model behaves as
generalized Chaplygin gas model and for $\eta\rightarrow -1$, the GCCG acts like de Sitter fluid. Exploiting (\ref{DE}), the
solution of energy density for GCCG is derived as \cite{62}
\begin{equation}
\rho_{d}=\left[K +
\left(1+C_2~(1+z)^{3(1+\mu)(1+\eta)}\right)^{\frac{1}{1+\eta}}\right]^{\frac{1}{1+\mu}}
\end{equation}
where $C_1$ is constant. This expression is written as
\begin{widetext}
\begin{equation}\label{rhod2}
\rho_{d}=\rho_{_{d0}}\left[C_s +(1-C_s)
\left(B_{s}+(1-B_{s})(1+z)^{3(1+\mu)(1+\eta)}\right)^{\frac{1}{1+\eta}}\right]^{\frac{1}{1+\mu}}
\end{equation}
where $z$ is the redshift, $B_{s}=\frac{1}{1+C_2}$,
$A_s=\left(1+K^{-1}B_{s}^{-\frac{1}{1+\eta}}\right)^{-1}$ and
$\rho_{d0}^{1+\mu}=K+B_{s}^{-\frac{1}{1+\eta}}$.\\
So from (\ref{E}), (\ref{Om}) and (\ref{rhod2}), we obtain
\begin{eqnarray}
H^2(z)= && H_0^2\left[\Omega_{m0}(1+z)^3+\Omega_{k0}(1+z)^2
+\Omega_{\Lambda 0}+\frac{\Omega_{k0}^2(1+z)^4}{4\Omega_{\Lambda
0}} \right.  \nonumber
\\
&&+\left. \left(1-\Omega_{m0}-\Omega_{k0}-\Omega_{\Lambda
0}-\frac{\Omega_{k0}^2}{4\Omega_{\Lambda 0}}\right)\left[A_s
+(1-A_s)
\left(B_{s}+(1-B_{s})(1+z)^{3(1+\mu)(1+\eta)}\right)^{\frac{1}{1+\eta}}\right]^{\frac{1}{1+\mu}}\right]
\nonumber\\
\end{eqnarray}
\end{widetext}
\newpage
\section{Methodology}\label{Sec3}
\begin{figure*}[htbp]
\centering
\includegraphics[scale=0.45]{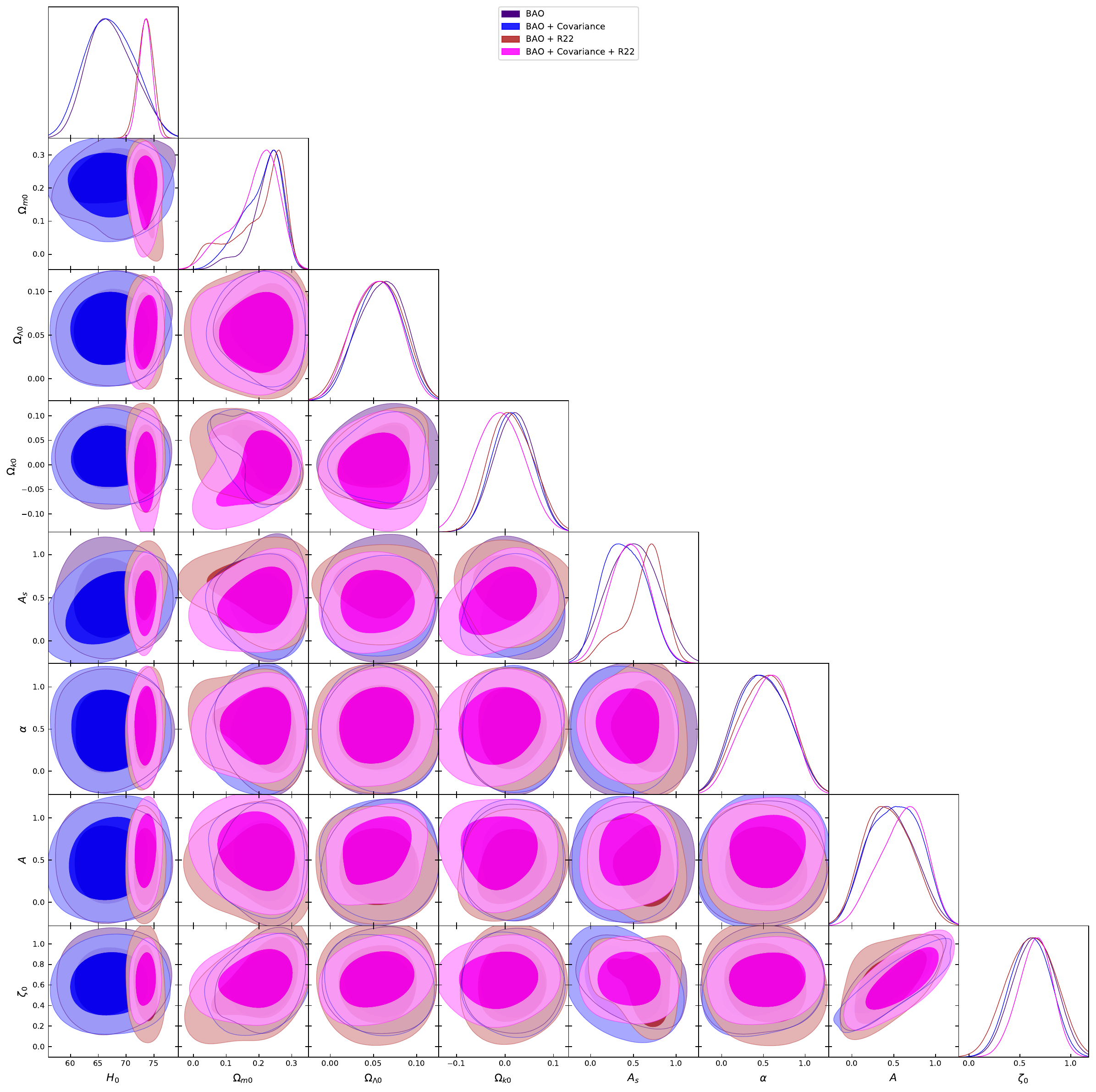}
\caption{The posterior distributions for a VMMG model, with and without a randomly generated covariance matrix, show negligible differences in distributions when considering datasets with between zero and fourteen components.}\label{fig_0}
\end{figure*}

\begin{figure*}[htbp]
\centering
\includegraphics[scale=0.45]{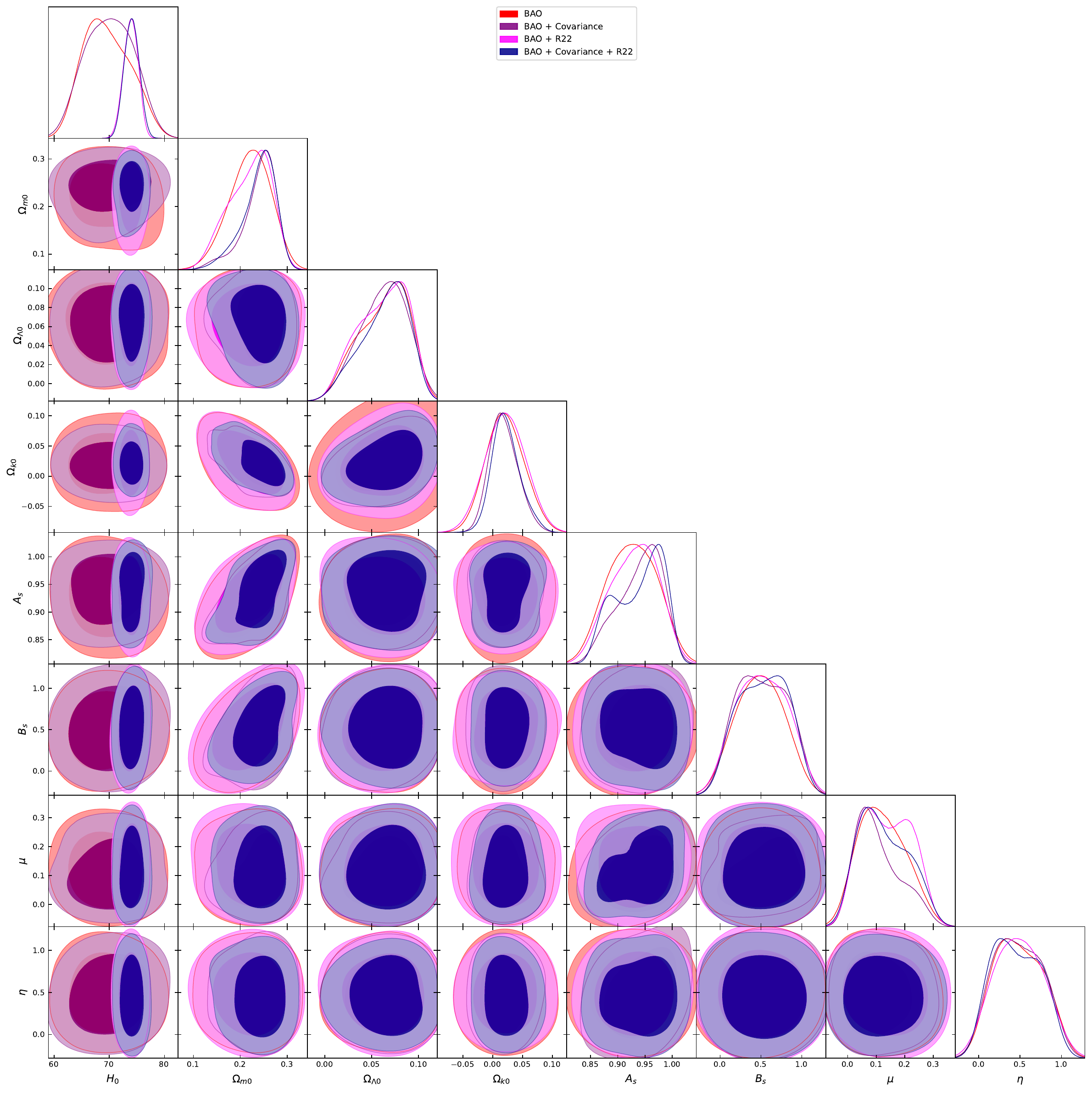}
\caption{The posterior distributions for a GCCG model, with and without a randomly generated covariance matrix, show negligible differences in distributions when considering datasets with between zero and fourteen components.}\label{fig_0.2}
\end{figure*}

In our investigation, we meticulously curated recent observations of Baryon Acoustic Oscillations (BAOs) sourced from a variety of galaxy surveys, with particular emphasis on leveraging data from the Sloan Digital Sky Survey (SDSS) \cite{42BAO,43BAO,44BAO,45BAO,46BAO,47BAO}. Additionally, we incorporated datasets from other reputable sources such as the Dark Energy Survey (DES) \cite{48BAO}, Dark Energy Camera Legacy Survey (DECaLS) \cite{49BAO}, and the 6dFGS Baryon Acoustic Oscillation survey (6dFGS BAO) \cite{50BAO} to ensure the breadth and diversity of our dataset. Recognizing the potential interdependencies among our selected data points, we diligently addressed this concern. While our primary aim was to mitigate highly correlated points, we also recognized the necessity of managing any existing correlations to uphold the integrity and precision of our analysis. To accurately estimate systematic errors, we employed mock data generated through N-body simulations. These simulations helped us determine covariance matrices precisely, which are crucial for understanding how different data points relate to each other. However, since measurements from various observational surveys differ in nature, obtaining accurate covariance matrices posed a significant challenge. To tackle this challenge, we followed an approach outlined in previous studies. We used a simple form where the diagonal elements of the covariance matrix ($C_{ii}$) were set equal to the square of the 1$\sigma$ errors ($\sigma_i^2$). This method, described in earlier research, helped us handle the complexities of covariance estimation. To replicate interdependencies within our chosen subset, we incorporated off-diagonal components into the covariance matrix, maintaining symmetry. This involved randomly selecting pairs of data points and assigning non-diagonal elements based on the product of the 1$\sigma$ errors of the respective data points. By incorporating non-negative correlations in this manner, we effectively represented correlations within 55\% of our chosen BAO dataset. Visual representations of our analysis, shown in Fig \ref{fig_0} and Fig \ref{fig_0.2}, illustrate the posterior distributions for the VMMG and GCCG Models. We compared results obtained with and without a test random covariance matrix containing fourteen components. Surprisingly, we found that the impact of the covariance matrix on our results was minimal, even when we varied the number of components. This finding underscores the robustness of our analysis methodology. In our ongoing research, we expanded our BAO dataset by integrating thirty independent measurements of the Hubble parameter sourced from the cosmic chronometers (CC) technique, as referenced in prior studies \cite{CC1,CC2,CC3,CC4}. Additionally, we augmented our dataset with the most recent Pantheon sample data focusing on Type Ia Supernovae (SNIa) \cite{SNIa}. To enrich our investigation, we introduced supplementary datasets, including 24 binned measurements of quasar (Q) distance modulus \cite{Quasar} and a comprehensive set of 162 Gamma-Ray Bursts (GRBs) \cite{GRB}, both of which have been extensively discussed in previous literature. Furthermore, to enhance the robustness of our analysis, we incorporated the latest measurement of the Hubble constant (R22) as an additional prior constraint \cite{7BAO}. In our analysis, we employed a nested sampling technique using the Polychord package, cited as \cite{51BAO}. This approach efficiently explores high-dimensional parameter spaces, allowing us to obtain robust results. Additionally, we utilized the GetDist package \cite{Getdist}, to present our findings in a clear and informative manner. GetDist facilitates the visualization and interpretation of our results, enhancing their accessibility to both researchers and broader audiences.\\\\
%%%%%%%%%%%%%%%%%%%%%%%%%%%%%%%%%%%%%%%%%%%%%%%%%%%%%%%%
%%%%%%%%%%%%%%%%%%%%%%%%%%%%%%%%%%%%%%%%%%%%%%%%%%%%%%%%

%%%%%%%%%%%%%%%%%%%%%%%%%%%%%%%%%%%%%%%%%%%%%%%%%%%%%%
\begin{figure*}[htbp]
\centering
\includegraphics[scale=0.45]{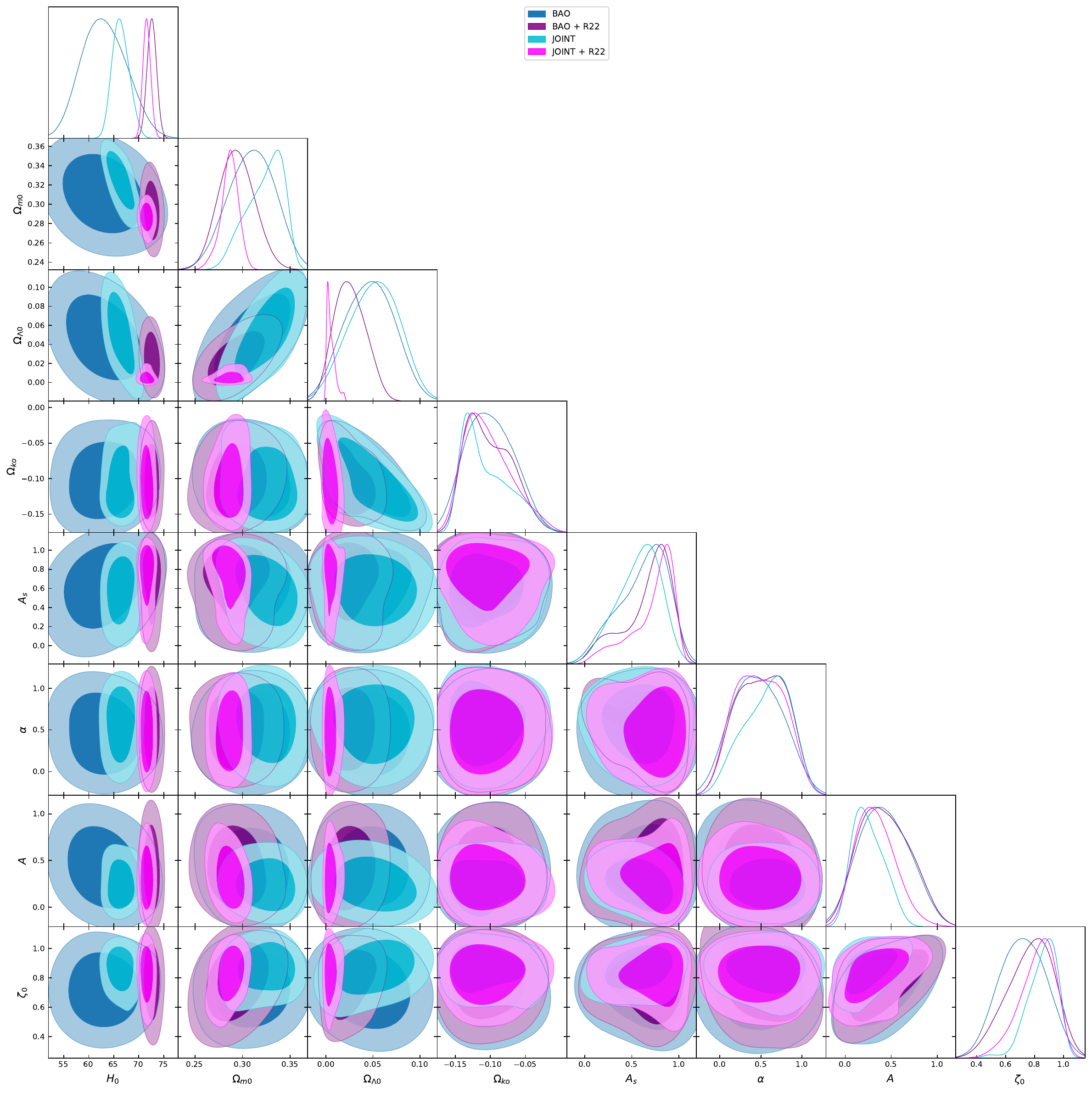}
\caption{The posterior distribution representing various cosmological parameters within the VMMG Model, with 1$\sigma$ and 2$\sigma$ confidence intervals.}\label{fig_2}
\end{figure*}
%%%%%%%%%%%%%%%%%%%%%%%%%%%%%%%%%%%%%%%%%%%%%%%%%%%%%%
\begin{figure*}[htbp]
\centering
\includegraphics[scale=0.45]{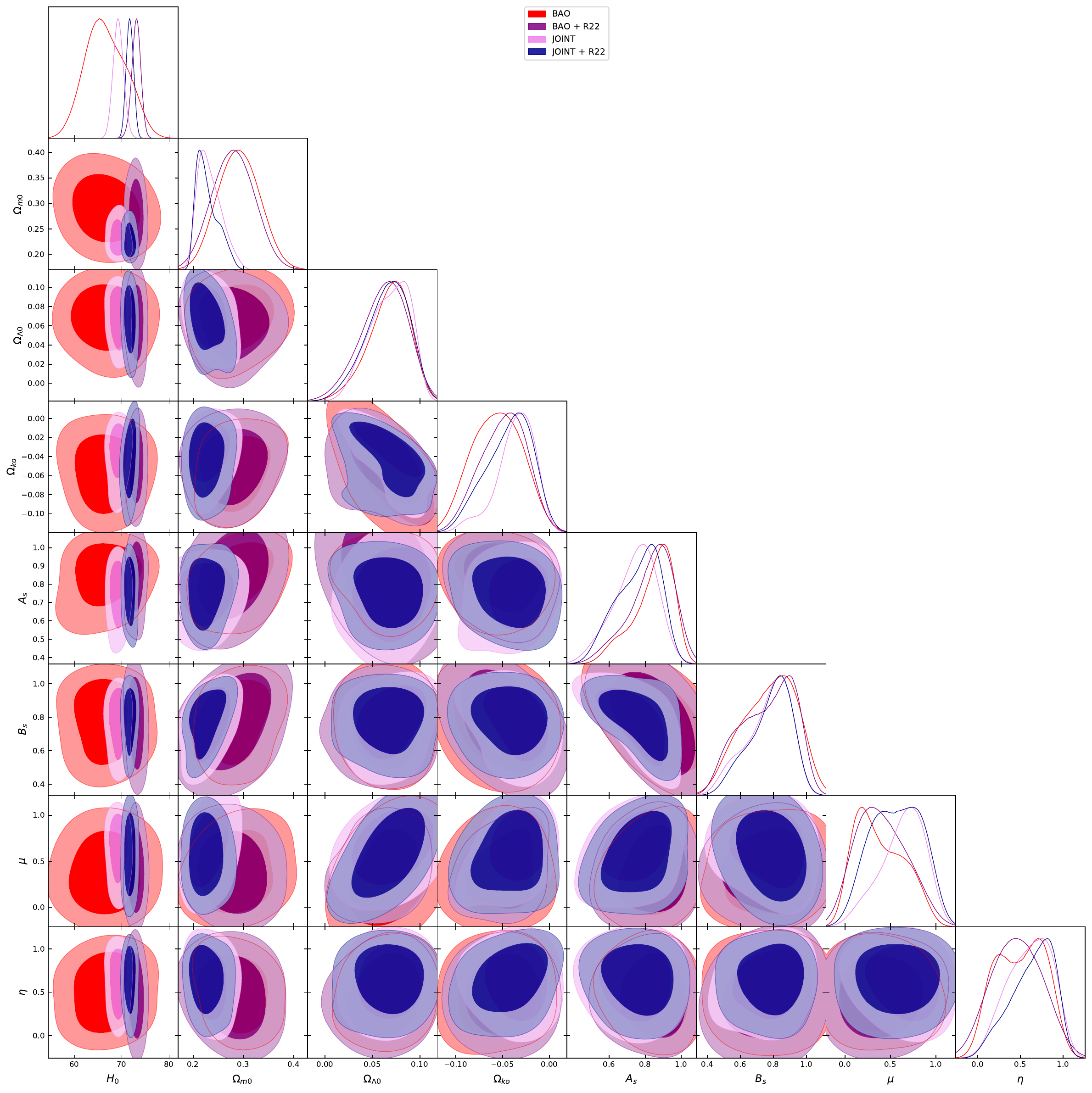}
\caption{The posterior distribution representing various cosmological parameters within the GCCG Model, with 1$\sigma$ and 2$\sigma$ confidence intervals.}\label{fig_3}
\end{figure*}
%%%%%%%%%%%%%%%%%%%%%%%%%%%%%%%%%%%%%%%%%%%%%%%%%
\begin{table*}
\centering
\begin{tabular}{|c|c|c|c|c|c|c|}
\hline
\multicolumn{7}{|c|}{MCMC Results} \\
\hline\hline
Model & Parameter & Priors & BAO & BAO + R22 & Joint & Joint + R22 \\[1ex]
\hline
& $H_0$ & [50,100] & $70.256500_{\pm 5.502574}^{\pm 9.206346}$ & $73.798696_{\pm 1.315504}^{\pm 2.543240}$ & $69.837362_{\pm 1.204873}^{\pm 2.478047}$ & $71.508817_{\pm 0.882681}^{\pm 1.614354}$ \\[1ex]
$\Lambda$CDM Model &$\Omega_{m0}$ &[0.,1.]  & $0.270161_{\pm 0.013827}^{\pm 0.033280}$ & $0.268255_{\pm 0.016790}^{\pm 0.043787}$ & $0.274526_{\pm 0.009858}^{\pm 0.021367}$ & $0.270609_{\pm 0.009678}^{\pm 0.022243}$ \\[1ex]
&$\Omega_{d0}$  & [0.,1.] & $0.725842_{\pm 0.011842}^{\pm 0.019262}$ & $0.727203_{\pm 0.013626}^{\pm 0.024379}$ & $0.721607_{\pm 0.007421}^{\pm 0.015585}$ & $0.724854_{\pm 0.007274}^{\pm 0.016354}$ \\[1ex]
&$r_{d}$(Mpc)  & [100.,200.] & $145.807497_{\pm 9.060325}^{\pm 13.994434}$ & $138.345762_{\pm 2.451777}^{\pm 4.939114}$ & $145.811932_{\pm 2.347736}^{\pm 4.545176}$ & $142.591327_{\pm 1.850175}^{\pm 3.951717}$ \\[1ex]
&$r_{d}/r_{fid}$  & [0.9,1.1] & $0.970661_{\pm 0.055989}^{\pm 0.068973}$ & $0.930500_{\pm 0.021272}^{\pm 0.028238}$ & $0.974064_{\pm 0.033510}^{\pm 0.057975}$ & $0.947662_{\pm 0.028335}^{\pm 0.043021}$ \\[1ex]
\hline
& $H_0$ & [50,100] & $63.076981_{\pm 4.476269}^{\pm 6.920660}$ & $72.623184_{\pm 0.877409}^{\pm 1.918024}$ & $66.397928_{\pm 1.628310}^{\pm 2.808849}$ & $71.612274_{\pm 0.675375}^{\pm 1.383053}$ \\[1ex]
& $\Omega_{m0}$ & [0.,1.] & $0.309681_{\pm 0.023961}^{\pm 0.044682}$ & $0.293673_{\pm 0.018701}^{\pm 0.029553}$ & $0.323977_{\pm 0.021140}^{\pm 0.039638}$ & $0.286455_{\pm 0.007862}^{\pm 0.021394}$ \\[1ex]
& $\Omega_{\Lambda0}$ & [0.,0.1] & $0.046372_{\pm 0.031249}^{\pm 0.041255}$ & $0.025434_{\pm 0.017776}^{\pm 0.024214}$ & $0.051223_{\pm 0.028907}^{\pm 0.047786}$ & $0.005226_{\pm 0.004198}^{\pm 0.005022}$ \\[1ex]
& $\Omega_{ko}$ & [-0.1,0.1] & $-0.100604_{\pm 0.038954}^{\pm 0.047937}$ & $-0.102529_{\pm 0.034833}^{\pm 0.043818}$ & $-0.106191_{\pm 0.033828}^{\pm 0.042359}$ & $-0.103724_{\pm 0.034139}^{\pm 0.044595}$ \\[1ex]
VMMG Model & $A_{s}$ & [0.,1.] & $0.623191_{\pm 0.316326}^{\pm 0.575108}$ & $0.669894_{\pm 0.300692}^{\pm 0.570953}$ & $0.578927_{\pm 0.246275}^{\pm 0.471071}$ & $0.723788_{\pm 0.250656}^{\pm 0.603990}$ \\[1ex]
& $\alpha$ & [0.,1.] & $0.456719_{\pm 0.350213}^{\pm 0.452516}$ & $0.510590_{\pm 0.364617}^{\pm 0.487738}$ & $0.577261_{\pm 0.333066}^{\pm 0.541369}$ & $0.482360_{\pm 0.349962}^{\pm 0.461090}$ \\[1ex]
& $A$ & [0.,1.] & $0.424731_{\pm 0.303642}^{\pm 0.409057}$ & $0.424469_{\pm 0.306861}^{\pm 0.389732}$ & $0.249583_{\pm 0.174369}^{\pm 0.242093}$ & $0.322933_{\pm 0.232177}^{\pm 0.301106}$ \\[1ex]
& $\zeta_{0}$ & [0.,1.] & $0.719277_{\pm 0.160434}^{\pm 0.281126}$ & $0.771839_{\pm 0.173162}^{\pm 0.339331}$ & $0.856714_{\pm 0.108135}^{\pm 0.205339}$ & $0.821935_{\pm 0.123697}^{\pm 0.283862}$ \\[1ex]
& $r_{d}$ (Mpc) & [100.,200.] & $151.444156_{\pm 9.740859}^{\pm 16.240184}$ & $135.122570_{\pm 2.867547}^{\pm 5.514129}$ & $145.962162_{\pm 2.284883}^{\pm 5.109759}$ & $141.546739_{\pm 1.527315}^{\pm 3.055590}$ \\[1ex]
& $r_{d}/r_{fid}$ & [0.9,1.1] & $1.002605_{\pm 0.065483}^{\pm 0.093906}$ & $0.919299_{\pm 0.014509}^{\pm 0.017935}$ & $0.974324_{\pm 0.027915}^{\pm 0.059446}$ & $0.947252_{\pm 0.023070}^{\pm 0.041848}$ \\[1ex]
\hline
& $H_0$ & [50,100] & $66.590458_{\pm 3.830490}^{\pm 7.943618}$ & $73.043659_{\pm 0.900965}^{\pm 1.921783}$ & $69.234519_{\pm 1.043849}^{\pm 2.085606}$ & $71.688017_{\pm 0.737745}^{\pm 1.386952}$ \\[1ex]
& $\Omega_{m0}$ & [0.,1.] & $0.290474_{\pm 0.044568}^{\pm 0.071365}$ & $0.279744_{\pm 0.045824}^{\pm 0.074140}$ & $0.232374_{\pm 0.024797}^{\pm 0.031183}$ & $0.227590_{\pm 0.022533}^{\pm 0.026817}$ \\[1ex]
& $\Omega_{\Lambda0}$ & [0.,0.1] & $0.067840_{\pm 0.020559}^{\pm 0.051789}$ & $0.062527_{\pm 0.025203}^{\pm 0.055079}$ & $0.068477_{\pm 0.023960}^{\pm 0.041632}$ & $0.066024_{\pm 0.023199}^{\pm 0.050106}$ \\[1ex]
& $\Omega_{ko}$ & [-0.1,0.1] & $-0.055808_{\pm 0.031984}^{\pm 0.041222}$ & $-0.048914_{\pm 0.027088}^{\pm 0.047402}$ & $-0.035985_{\pm 0.016567}^{\pm 0.054157}$ & $-0.042903_{\pm 0.026621}^{\pm 0.051167}$ \\[1ex]
GCCG Model& $A_{s}$ & [0.5,1.] & $0.845407_{\pm 0.115460}^{\pm 0.265680}$ & $0.835317_{\pm 0.129186}^{\pm 0.284070}$ & $0.746478_{\pm 0.125267}^{\pm 0.267879}$ & $0.762105_{\pm 0.138390}^{\pm 0.259584}$ \\[1ex]
& $B_{s}$ & [0.3,1.] & $0.777478_{\pm 0.195466}^{\pm 0.302082}$ & $0.767995_{\pm 0.189570}^{\pm 0.308703}$ & $0.770860_{\pm 0.150548}^{\pm 0.298653}$ & $0.784111_{\pm 0.151517}^{\pm 0.287660}$ \\[1ex]
& $\mu$ & [0.,1.] & $0.377164_{\pm 0.295752}^{\pm 0.350000}$ & $0.398506_{\pm 0.304323}^{\pm 0.380732}$ & $0.638717_{\pm 0.272800}^{\pm 0.541761}$ & $0.567148_{\pm 0.299906}^{\pm 0.506744}$ \\[1ex]
& $\eta$ & [0.,1.] & $0.502387_{\pm 0.340965}^{\pm 0.449156}$ & $0.461452_{\pm 0.320349}^{\pm 0.450812}$ & $0.616084_{\pm 0.289631}^{\pm 0.508848}$ & $0.644756_{\pm 0.266663}^{\pm 0.544710}$ \\[1ex]
& $r_{d}$ (Mpc) & [100.,200.] & $151.484609_{\pm 10.451305}^{\pm 16.504765}$ & $138.032959_{\pm 2.136997}^{\pm 3.975253}$ & $147.187439_{\pm 1.806239}^{\pm 4.771847}$ & $142.329222_{\pm 1.514611}^{\pm 2.801019}$ \\[1ex]
& $r_{d}/r_{fid}$ & [0.9,1.1] & $1.006738_{\pm 0.065848}^{\pm 0.095708}$ & $0.930724_{\pm 0.022444}^{\pm 0.028873}$ & $0.979584_{\pm 0.032241}^{\pm 0.059617}$ & $0.945289_{\pm 0.024865}^{\pm 0.039088}$ \\[1ex]
\hline
\end{tabular}
\caption{Constraints at the 95\% confidence level (CL) on the cosmological parameters for the standard $\Lambda$CDM, VMMG and GCCG models.}
\label{tab_2}
\end{table*}
%%%%%%%%%%%%%%%%%%%%%%%%%%%%%%%%%%%%%%%%%%%%%%%%%
\begin{figure*}[htb]
\begin{subfigure}{.32\textwidth}
\includegraphics[width=\linewidth]{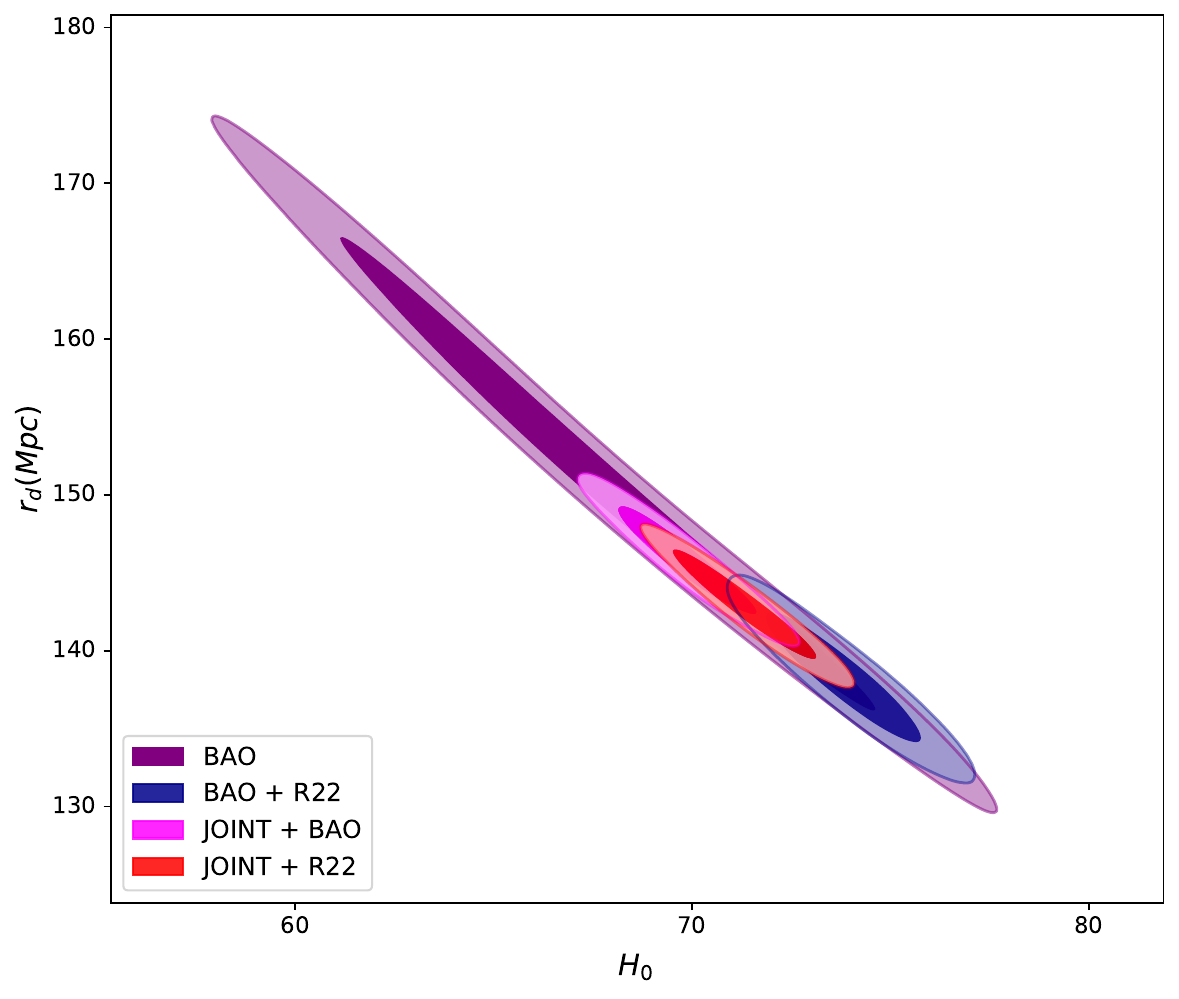}% width
    \caption{$\Lambda$CDM Model}
    \label{fig_4a}
\end{subfigure}
\hfil
\begin{subfigure}{.32\textwidth}
\includegraphics[width=\linewidth]{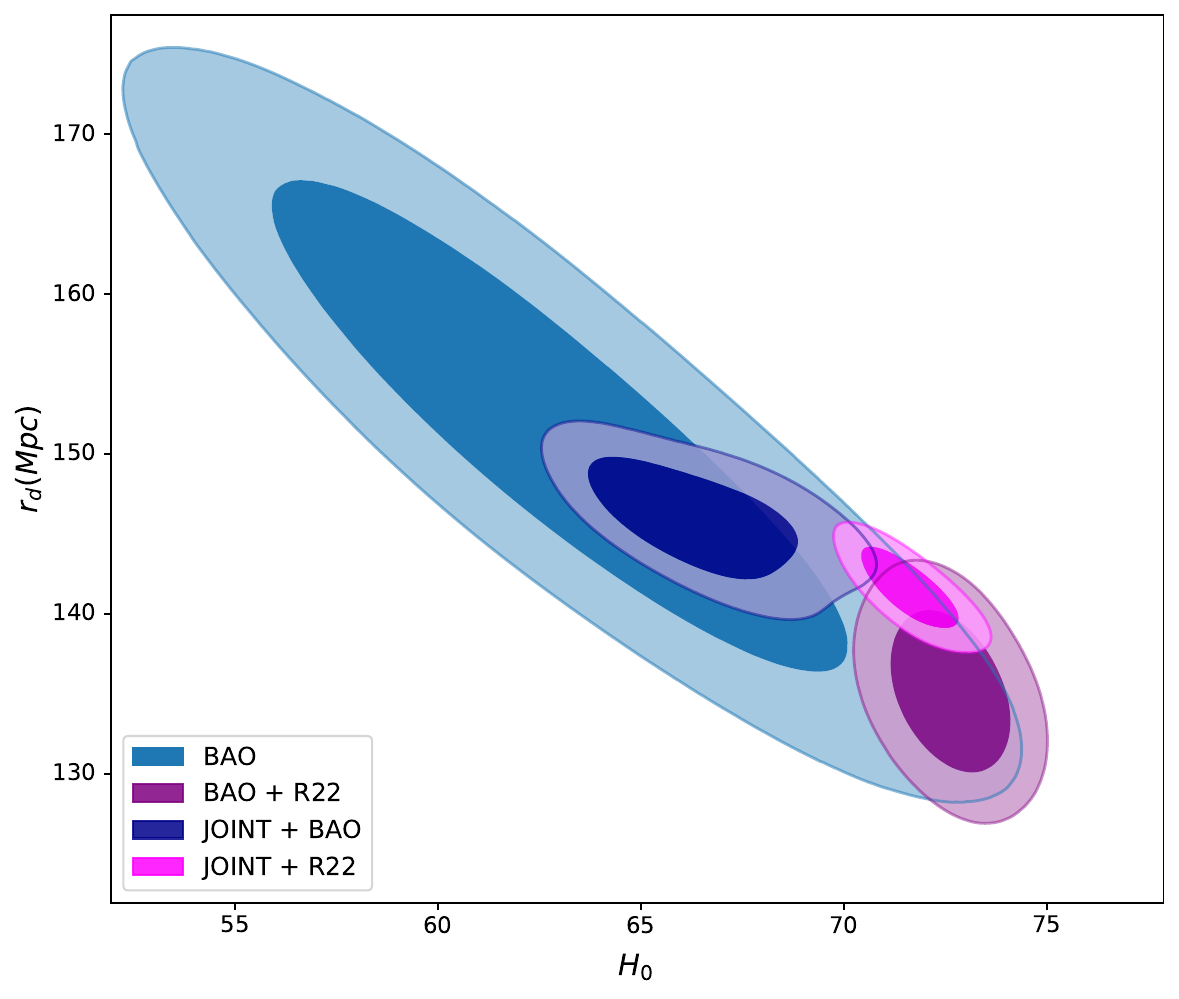}% width
    \caption{VMMG Model }
    \label{fig_4b}
\end{subfigure}
\hfil
\begin{subfigure}{.32\textwidth}
\includegraphics[width=\linewidth]{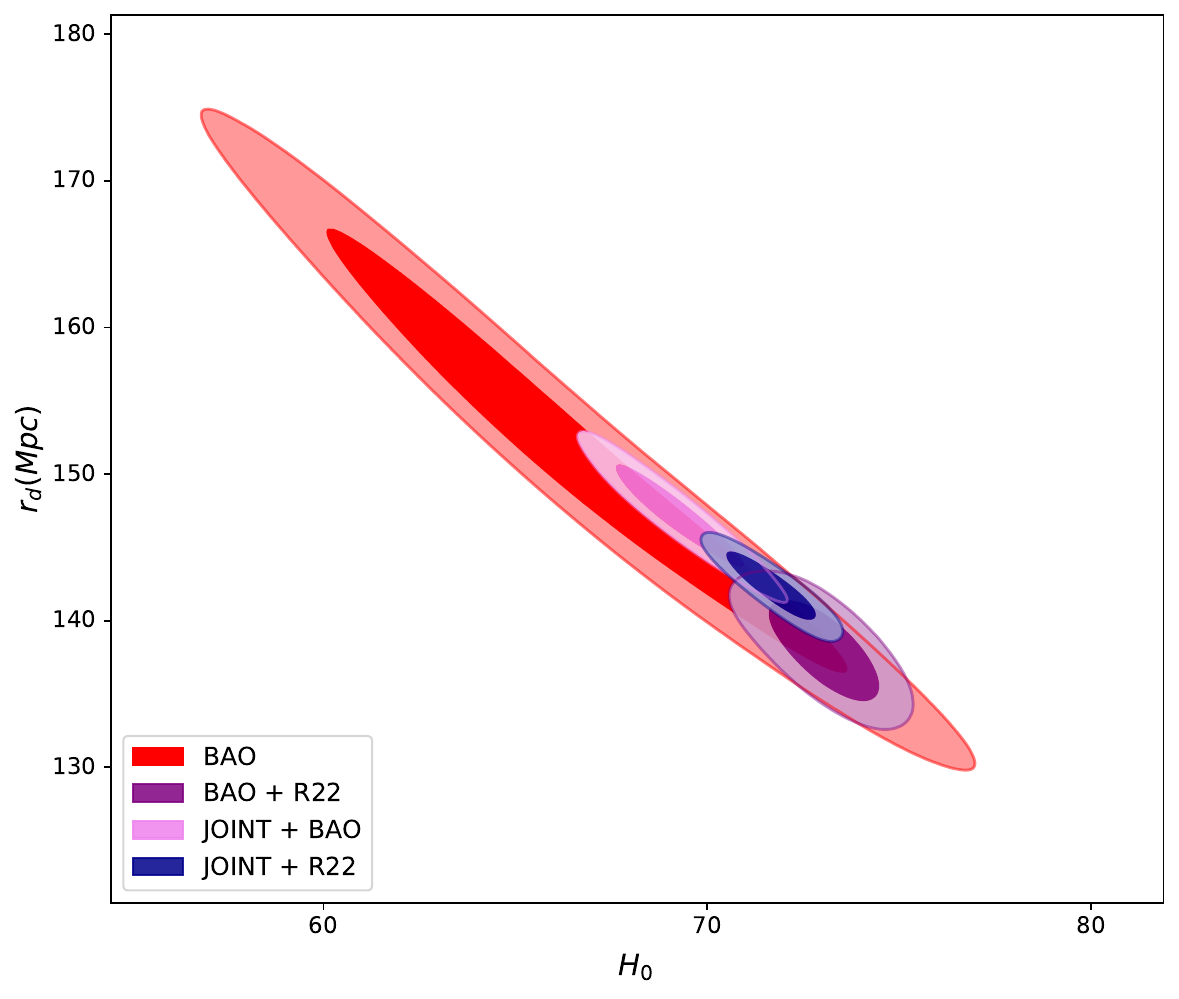}% width
    \caption{GCCG Model}
     \label{fig_4c}
\end{subfigure}
\caption{The posterior distributions of varied observational data within the $r_{d}$ vs $H_{0}$ contour plane are depicted for the $\Lambda$CDM, VMMG, and GCCG models. Shaded regions represent 1$\sigma$ and 2$\sigma$ confidence levels.}
\end{figure*}
%%%%%%%%%%%%%%%%%%%%%%%%%%%%%%%%%%%%%%%%%%%%%%%%%
\section{Cosmography Parameters}\label{Sec4}
Cosmography Parameters play a crucial role in our comprehension of the universe's dynamics and evolution. These parameters provide us with insights into the underlying structure, expansion, and behavior of the cosmos. In astrophysical research, they serve as fundamental tools for analyzing various cosmological models and understanding the intricate interplay between different components of the Universe \cite{weinberg1972gravitation}. One important aspect of Cosmography Parameters is their ability to shed light on the acceleration or deceleration of cosmic expansion. The deceleration parameter, often denoted as 'q', quantifies the rate at which the universe's expansion is slowing down. It is defined by the formula: \( q = -\frac{\ddot{a}a}{\dot{a}^2} \) where 'a' represents the scale factor of the universe, and dots denote derivatives with respect to time. A negative value of 'q' indicates an accelerating expansion, while a positive value suggests a decelerating one. This parameter holds significant implications for theories of dark energy and the ultimate fate of the Universe \cite{sahni2000case}. In addition to deceleration, higher-order parameters such as jerk and snap provide further insights into the dynamics of cosmic expansion. The jerk parameter, denoted by 'j', represents the rate of change of acceleration with time and is given by: \( j = \frac{\dddot{a}a}{\dot{a}^3} \). It offers a deeper understanding of the transitions in the universe's expansion rate, potentially revealing clues about the nature of dark energy and gravitational interactions at large scales \cite{visser2004jerk}. The snap parameter, denoted as 's', captures the rate of change of jerk with time and is expressed as: \( s = \frac{\ddddot{a}a}{\dot{a}^4} \). While the snap parameter is less commonly discussed compared to deceleration and jerk, it offers valuable insights into the finer details of cosmic dynamics, especially in scenarios where precise measurements are crucial for testing theoretical predictions \cite{visser2005cosmography}. These parameters collectively provide a comprehensive framework for characterizing the evolution of the universe and investigating the underlying physics governing its behavior.\\\\
%%%%%%%%%%%%%%%%%%%%%%%%%%%%%%%%%%%%%%%%%%%%%%%%%
\begin{figure*}[htb]
\begin{subfigure}{.32\textwidth}
\includegraphics[width=\linewidth]{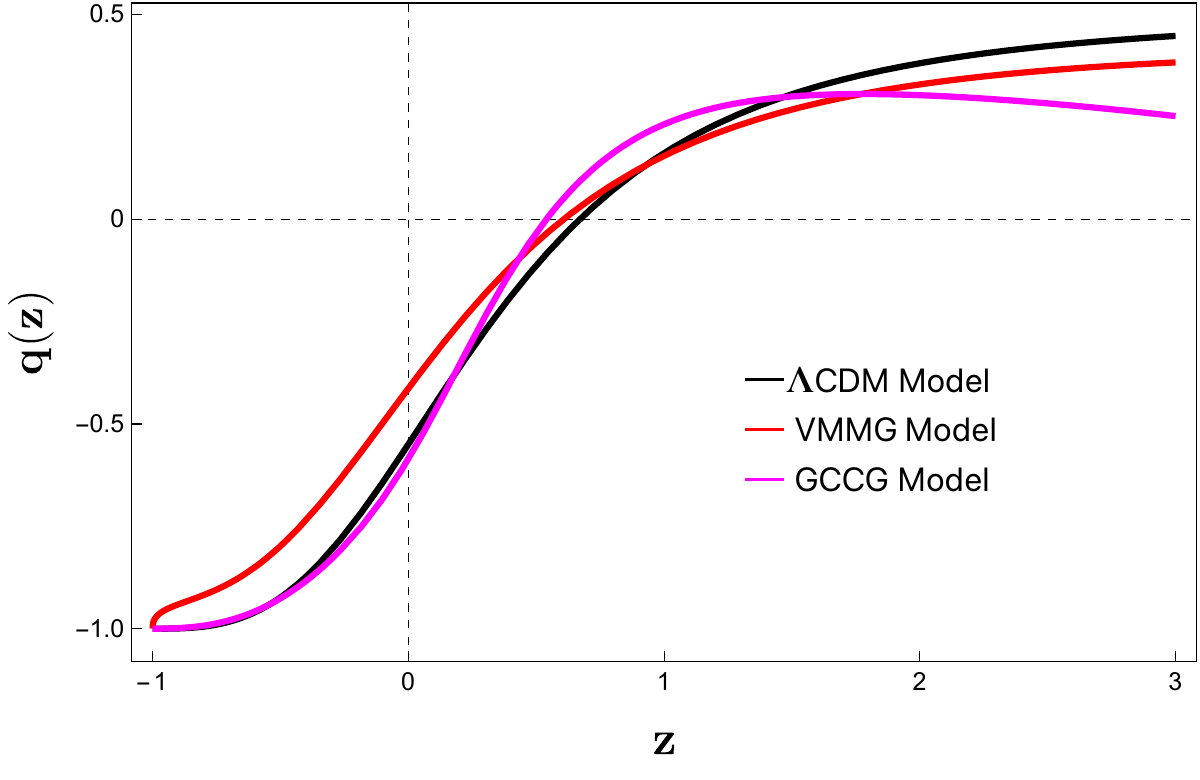}% width
    \caption{deceleration parameter}
    \label{fig_5a}
\end{subfigure}
\hfil
\begin{subfigure}{.32\textwidth}
\includegraphics[width=\linewidth]{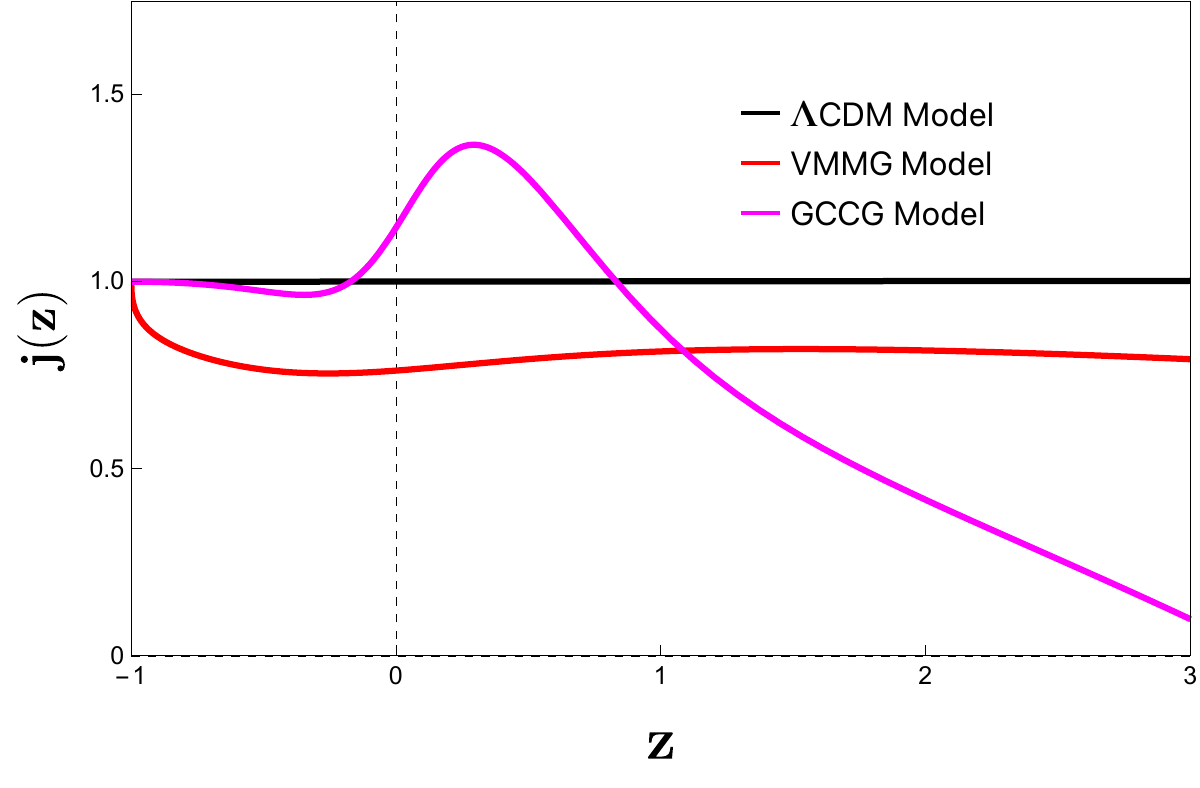}% width
    \caption{Jerk Parameter}
    \label{fig_5b}
\end{subfigure}
\hfil
\begin{subfigure}{.32\textwidth}
\includegraphics[width=\linewidth]{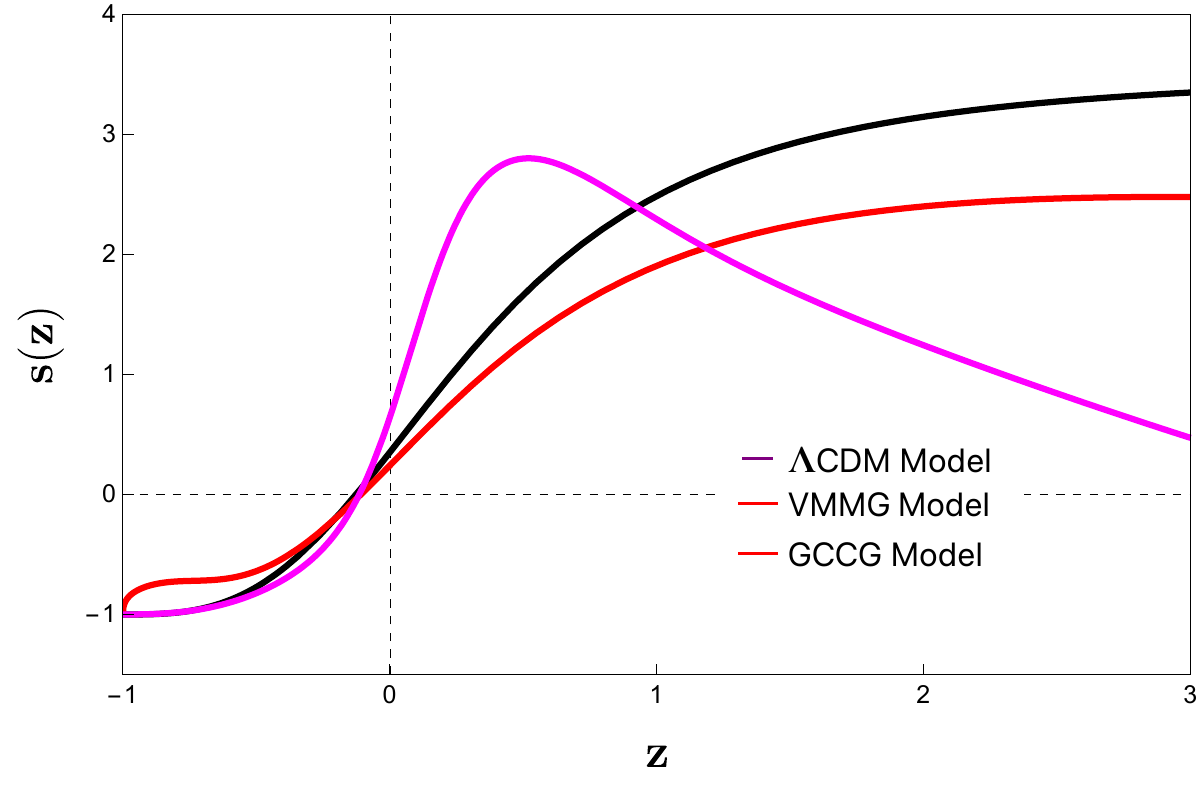}% width
    \caption{Snap Parameter}
     \label{fig_5c}
\end{subfigure}
\caption{The evolution of various Cosmography parameters with respect to redshift ($z$). It includes the $\Lambda$CDM model shown by the black line, the VMMG model depicted by the red line, and the GCCG model represented by the magenta line, utilizing the best-fit values obtained from the Joint analysis.}\label{fig_5}
\end{figure*}
%%%%%%%%%%%%%%%%%%%%%%%%%%%%%%%%%%%%%%%%%%%%%%%%%%%%%%%%%%
\begin{table*}
\begin{tabular}{|c|c|c|c|c|c|c|c|}
\hline
\multicolumn{8}{|c|}{Numerical values of different cosmography parameters at different epochs.} \\
\hline\hline
Models & Redshift & deceleration epochs & Numerical Value & jerk epochs & Numerical Value & snap epochs & Numerical Value \\
\hline
& $z \longrightarrow \infty$ & $\left(q_{i}\right)$ & 0.449061 & $\left(j_{i}\right)$ & 1 & $\left(s_{i}\right)$ & 3.359387  \\

$\Lambda$CDM Model & $z \longrightarrow 0$ & $\left(q_{0}\right)$ & -0.550623 & $\left(j_{0}\right)$ & 1 & $\left(s_{0}\right)$ & 0.357323 \\

& $z \longrightarrow-1$ & $\left(q_{f}\right)$ & -1 & $\left(j_{f}\right)$ & 1 & $\left(s_{f}\right)$ & -1 \\

& $z_{t r}$ & $(q=0)$ & 0.682484 & $-$ & $-$ & $(s=0)$ & -0.118272\\
\hline

& $z \longrightarrow \infty$ & $\left(q_{i}\right)$ & 0.382961 & $\left(j_{i}\right)$ & 0.795048 & $\left(s_{i}\right)$ & 2.483671  \\

VMMG Model & $z \longrightarrow 0$& $\left(q_{0}\right)$ & -0.417005 & $\left(j_{0}\right)$ & 0.764826 & $\left(s_{0}\right)$ & 0.246663 \\

& $z \longrightarrow-1$&$\left(q_{f}\right)$ & -1 & $\left(j_{f}\right)$ & 1 & $\left(s_{f}\right)$ & -1 \\

& $z_{t r}$& $(q=0)$ & 0.610979 & $-$ & $-$ & $(s=0)$ & -0.118272 \\
\hline

& $z \longrightarrow \infty$ & $\left(q_{i}\right)$ & 0.252849 & $\left(j_{i}\right)$ & 0.102984 & $\left(s_{i}\right)$ & 0.478429  \\

GCCG Model & $z \longrightarrow 0$ & $\left(q_{0}\right)$ & -0.585917 & $\left(j_{0}\right)$ & 1.137192 & $\left(s_{0}\right)$ & 0.617903 \\

& $z \longrightarrow-1$&$\left(q_{f}\right)$ & -1 & $\left(j_{f}\right)$ & 1 & $\left(s_{f}\right)$ & -1 \\

& $z_{t r}$& $(q=0)$ & 0.540476 & $-$ & $-$ & $(s=0)$ & -0.118272 \\
\hline
\end{tabular}
\caption{The values of the different Cosmography parameters at various epochs and the redshift corresponding to the phase transition are provided for the $\Lambda$CDM, VMMG, and GCCG models.}\label{tab_3}
\end{table*}
\section{Dark Energy Dynamics: Statefinder Diagnostics and the $O_{m}$ Test}\label{Sec5}
Understanding the nature and dynamics of Dark Energy (DE) in the cosmos is paramount to unraveling the mysteries of the Universe's evolution. To achieve this, cosmologists employ sophisticated diagnostic tools that offer insights into the behavior of DE models. Two such prominent methods are the statefinder diagnostics \cite{statefinder1,statefinder2,statefinder3,statefinder4} and the $O_{m}$ diagnostic test \cite{sahni2008two}, each offering unique perspectives on probing DE dynamics. The statefinder diagnostics formalism provides a powerful means to differentiate between various DE models and compare their behaviors. By utilizing higher-order derivatives of the scale factor, this method allows for a model-independent exploration of cosmic characteristics. The key diagnostic pair $\{r, s\}$, where $r$ and $s$ are dimensionless parameters, offers valuable insights into the nature of DE. In the $r-s$ plane, specific pairs correspond to standard DE models, such as $\{r,s\}=\{1,0\}$ representing the $\Lambda$CDM model and $\{r,s\}=\{1,1\}$ indicating the standard cold dark matter model (SCDM). Trajectories in this plane illustrate the time evolution of different DE models, with $s > 0$ and $s < 0$ defining quintessence-like and phantom-like models, respectively. Deviations from standard values signify distinct characteristics of DE models. Introduced as a robust diagnostic tool, the $O_{m}$ diagnostic test relies solely on directly measurable quantities, particularly the Hubble parameter $H(z)$ obtained from theoretical assumption. This simplicity enhances its practicality and reliability in discerning between cosmological scenarios. The diagnostic test serves to distinguish between cosmological models, with $O_{m} = \Omega_{m0}$ indicating consistency with the $\Lambda$CDM model. Deviations, where $O_{m} > \Omega_{m0}$ or $O_{m} < \Omega_{m0}$, suggest quintessence or phantom scenarios \cite{escamilla2016nonparametric}, respectively.\\\\
\begin{figure*}[htb]
\begin{subfigure}{.32\textwidth}
\includegraphics[width=\linewidth]{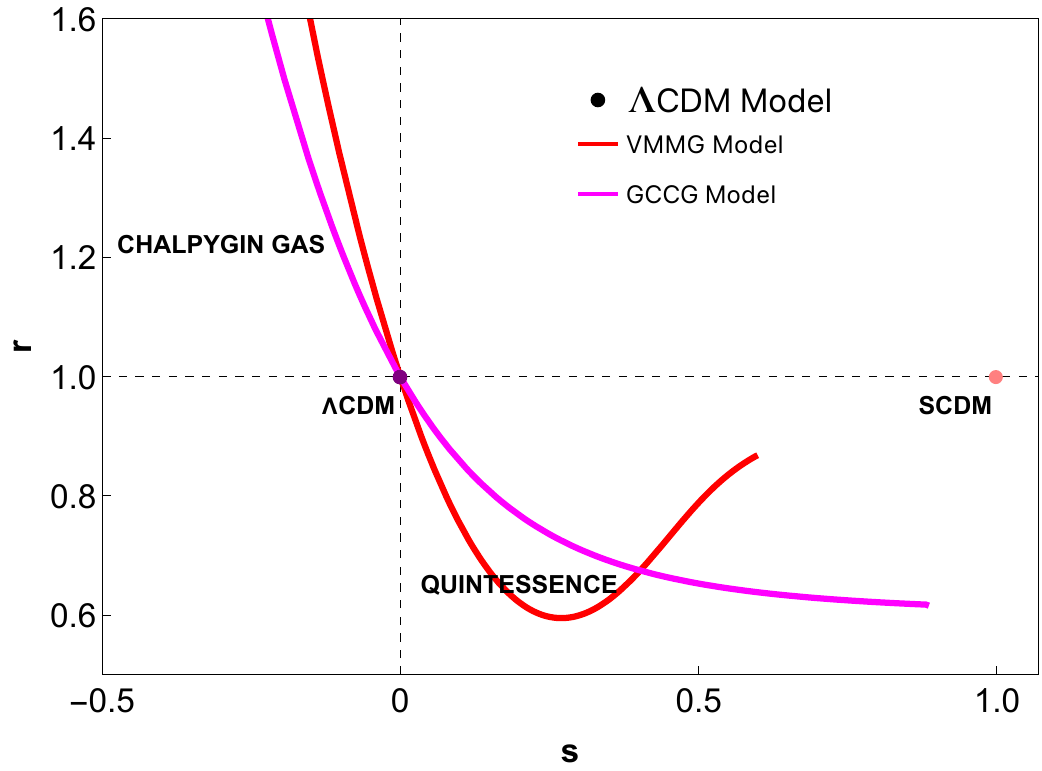}% width
    \caption{$\{s, r\}$ profile}
    \label{fig_6a}
\end{subfigure}
\hfil
\begin{subfigure}{.32\textwidth}
\includegraphics[width=\linewidth]{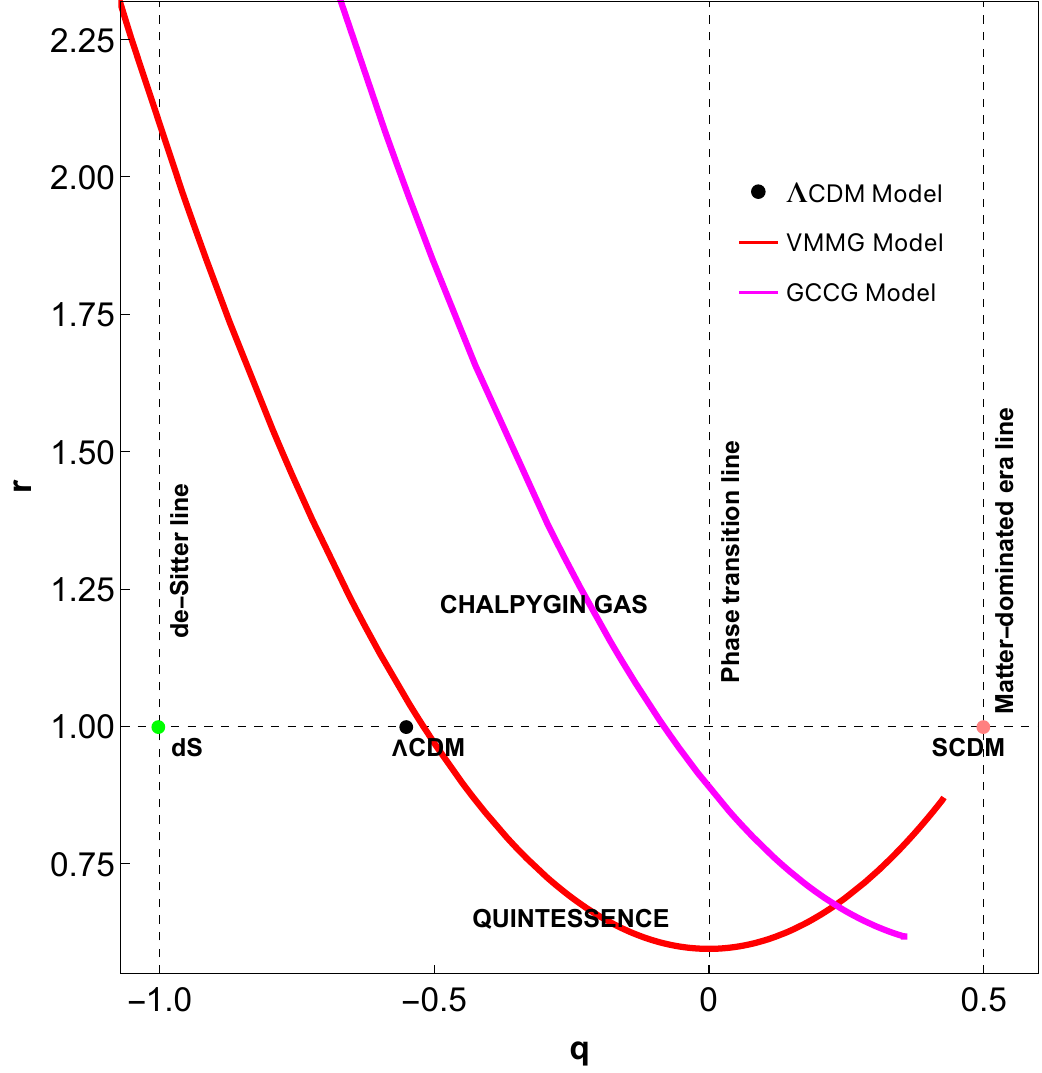}% width
    \caption{$\{q, r\}$ profile}
    \label{fig_6b}
\end{subfigure}
\hfil
\begin{subfigure}{.32\textwidth}
\includegraphics[width=\linewidth]{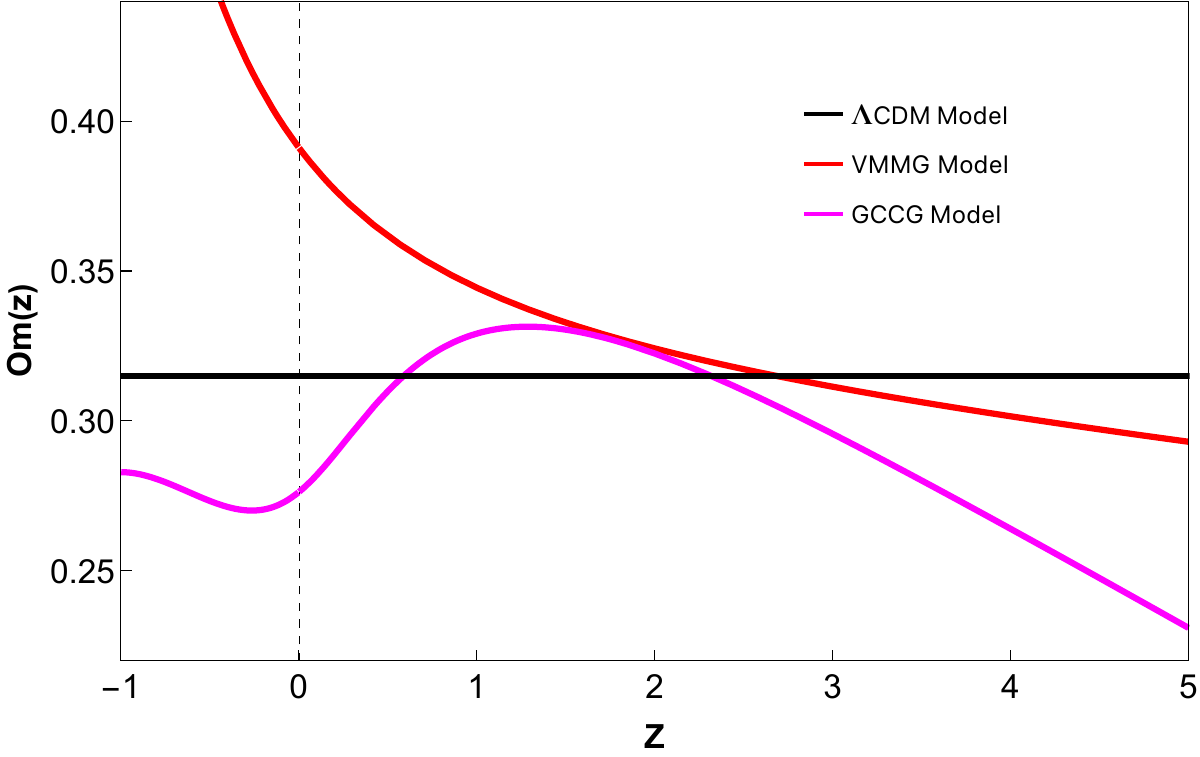}% width
    \caption{$O_{m}$ Diagnostic}
     \label{fig_6c}
\end{subfigure}
\caption{The evolution of the Statefinder diagnostic and the $O_{m}$ diagnostic profile. It includes the VMMG model represented by the red line and the GCCG model depicted by the magenta line, utilizing the best-fit values obtained from the joint analysis.}\label{fig_6}
\end{figure*}
%%%%%%%%%%%%%%%%%%%%%%%%%%%%%%%%%%%%%%%%%%%%%%%%%
\section{Results}\label{Sec6}
Fig \ref{fig_2} and \ref{fig_3} depict the posterior distributions of crucial cosmological parameters, with contours representing the \(68\%\) and \(95\%\) confidence levels for the VMMG and GCCG models, respectively. The detailed results obtained from the MCMC simulations are presented in Table \ref{tab_2}. When we include the R22 prior in the Joint dataset, the best-fitting value for \(H_{0}\) differs from the findings in \cite{1BAO} but aligns closely with measurements from the SNIe sample in \cite{7BAO}. Conversely, without R22 priors in the Joint dataset, the estimated \(H_{0}\) aligns more closely with the value reported in \cite{1BAO} across all three models. Notably, the optimal values for the matter density, denoted as \(\Omega_{m0}\), exhibit deviations: in the VMMG Model, they appear higher than those documented in \cite{1BAO}, whereas in the GCCCG Model, the optimal value of \(\Omega_{m0}\) is lower than reported in \cite{1BAO}. The BAO scale, a key aspect in cosmology, stems from the cosmic sound horizon imprinted in the CMB during the drag epoch (\(z_d\)), denoting baryon-photon decoupling. \(r_d\), the BAO scale, is determined by integrating \(c_s/H\) from \(z_d\) to infinity, where \(c_s\) depends on pressure perturbation in photons and energy density perturbations in baryons and photons. It simplifies to \(1/\sqrt{3(1+R)}\) with \(R = \delta \rho_B / \delta \rho_\gamma\). Observationally, \(z_d = 1059.94 \pm 0.30\), and in a $\Lambda$CDM model, \(r_d = 147.09 \pm 0.26\) Mpc according to \cite{1BAO}. In the exploration of the VMMG model, Fig~ \ref{fig_4b} illustrates the posterior distribution concerning the \(r_{d}-H_{0}\) contour plane. Initially focusing on the BAO datasets, the inferred value for \(r_{d}\) stands at \(151.444 \pm 9.740\) Mpc. However, when the R22 prior is integrated exclusively into the BAO dataset, it leads to a revised estimate of the sound horizon at the drag epoch, now at \(135.122 \pm 2.867\) Mpc. Moving on to the Joint dataset, the derived value for \(r_{d}\) is \(145.962 \pm 2.28\) Mpc, demonstrating a close alignment with the findings presented in \cite{1BAO}. Further refinement occurs with the incorporation of the R22 prior into the full dataset, resulting in \(r_{d}\) being estimated at \(141.546 \pm 1.527\) Mpc, which notably corresponds closely with the analysis in \cite{68BAO}. Similarly, within the framework of the GCCG Model, Fig~\ref{fig_4c} depicts the posterior distribution regarding the \(r_{d}-H_{0}\) contour plane. Initially, considering the BAO datasets in isolation, the estimated \(r_{d}\) stands at \(151.484 \pm 10.451\) Mpc. Upon integrating the R22 prior exclusively into the BAO dataset, the inferred \(r_{d}\) adjusts to \(138.032 \pm 2.136\) Mpc. Transitioning to the Joint dataset, \(r_{d}\) is estimated at \(147.187 \pm 1.806 \) Mpc, exhibiting a close match with the findings in \cite{1BAO}. Finally, with the inclusion of the R22 prior in the complete dataset, the estimate for \(r_{d}\) becomes \(142.329 \pm 1.514\) Mpc, indicating a notable agreement with the analysis presented in \cite{68BAO}. To conduct a comparative study between the different cosmography parameters in the $\Lambda$CDM, VMMG, and GCCG models, we have plotted the evolution of these parameters at various epochs, as shown in Figure \ref{fig_5}. In the figures, the redshift $z$ approaches infinity, corresponding to the early Universe, while the redshift $z$ approaches 0, corresponding to the late-time Universe. A redshift $z$ approaching -1 corresponds to the future Universe. Fig \ref{fig_5a} illustrates the evolution of the deceleration parameter for each models. In the case of the $\Lambda$CDM Model, the deceleration parameter starts from a positive value ($q_i$) in the early stages, becomes negative at the present ($q_0$), and approaches -1 in the future ($q_f$), corresponding to the de Sitter phase. This transition signifies the shift from decelerated expansion to accelerated expansion. A similar behavior can be observed in the VMMG and GCCG Models, mirroring the $\Lambda$CDM model, where transitions from positive to negative values indicate a similar evolution from deceleration to acceleration. Fig \ref{fig_5b} depicts the evolution of the jerk parameters for each model, starting with the $\Lambda$CDM Model. The jerk parameter remains constant at 1 for all epochs, indicating a consistent rate of change of acceleration. In the case of the VMMG Model, the value of the jerk parameter is slightly lower than that of the $\Lambda$CDM Model throughout the evolution until it approaches $z = -1$, where it matches the value of the $\Lambda$CDM Model. In the case of the VMMG Model, the value of the jerk is lower than the value predicted by the $\Lambda$CDM Model in the redshift range of \(0.826 < z < 3\). However, as the redshift decreases, one could observe the value of the jerk becoming higher than the value predicted by $\Lambda$CDM. Eventually, as it approaches $z = -1$, it aligns with the value predicted by the $\Lambda$CDM Model. Fig \ref{fig_5c} depicts the evolution of the snap parameters for each model, starting with the $\Lambda$CDM Model. The snap parameter initiates from a high positive value ($s_i$) at early times, decreases to a small positive value ($s_0$) at present, and approaches $-1$ in the future ($s_f$). It characterizes the deviation from simple exponential expansion. A similar behavior can be observed in the VMMG Model, where the snap parameter shows a comparable trend of decreasing from early times to the present and approaching $-1$ in the future. In the case of the GCCG Model, the snap parameter begins from a small positive value ($s_i$) at early times, increases to a higher positive value as the redshift decreases, and approaches $-1$ in the future ($s_f$). Overall, these models provide different theoretical frameworks for understanding the dynamics and evolution of the universe, with each model capturing specific aspects of cosmic behavior and expansion. The numerical values of different cosmography parameters at different redshift epochs are presented in Table \ref{tab_3}. Fig \ref{fig_6} illustrates different diagnostic parameters of the VMMG and GCCG models. In Fig \ref{fig_6a}, the evolution of the $\{s, r\}$ profile for both models is depicted. The red line represents the evolution of the VMMG Model, indicating its behavior. During the early stages, the model assumes values within the range $r < 1$ and $s > 0$, which corresponds to the quintessence region. This suggests that quintessence, a form of dark energy with a dynamic EoS, dominates the energy content in the early Universe. As the Universe progresses, the model reaches the fixed point $\{r, s\}=\{1, 0\}$, signifying a transition to the $\Lambda$CDM point depicted by the black line. In this late-time phase, $r$ remains greater than 1 while $s$ becomes negative. This behavior characterizes the Chaplygin gas region, which is a model incorporating a generalized form of dark energy. A similar behavior can be observed in the GCCG Model, represented by the magenta line. Fig \ref{fig_6b} illustrates the $\{q, r\}$ profile for both models. Initially, in the VMMG model, the parameters exhibit values where $q > 0$ and $r < 1$, representing the quintessence region. This indicates an accelerating expansion of the Universe, with the dominance of quintessence energy density over matter density, driving the accelerated expansion. As the evolution progresses, the model transitions to values where $q < 0$ and $r > 1$, corresponding to the Chaplygin gas region. Here, the negative $q$ denotes a decelerating expansion of the Universe, while $r > 1$ suggests that the dominance of Chaplygin gas over matter causes deceleration by crossing the Finally, the model tends towards the de Sitter point located at $(-1, 1)$. Similar behavior can be observed in the GCCG Models as well shown in magenta line. Fig. \ref{fig_6c} illustrates that for the VMMG model, the diagnostic value \(O_{m}\) is initially less than \(\Omega_{m0}\) at high redshifts, indicating that the VMMG model falls within the phantom region. As redshift decreases, \(O_{m}\) surpasses \(\Omega_{m0}\), suggesting a transition to the quintessence domain. Conversely, in the GCCG Model, \(O_{m}\) is less than \(\Omega_{m0}\) at high redshifts, indicating phantom behavior. However, it exhibits quintessence behavior for a smaller redshift range (\(0.67 < z < 2.32\)). Interestingly, at low redshifts, \(O_{m}\) again becomes smaller than \(\Omega_{m0}\), leading to the model falling back into the phantom region.nIn our exploration of cosmological models, we employ both the Akaike Information Criterion (AIC) and the Bayesian Information Criterion (BIC). AIC is computed using the maximum likelihood (\(\mathcal{L}_{\max}\)) without the R22 prior, and the number of parameters (\(k\)) against the total data points (\(N_{\text{tot}}\)). BIC incorporates \(\mathcal{L}_{\max}\), \(k\), and \(N_{\text{tot}}\). For \(\Lambda\)CDM, VMMG, and GCCG models, the AIC and BIC values respectively are \([277.38, 284.89, 282.89]\) and \([277.59, 285.73, 284.89]\). Despite \(\Lambda\)CDM showing the best fit, all models receive support from our AIC and BIC results, indicating none can be dismissed based on existing data. Evaluation of VMMG and GCCG models relative to \(\Lambda\)CDM involves a reduced chi-square statistic defined as $\chi_{\text{red}}^{2} = \chi^{2} / \text{Dof}$, with values \([0.961, 0.970, 0.968]\), close to 1, indicating satisfactory alignment with observed data.
%%%%%%%%%%%%%%%%%%%%%%%%%%%%%%%%%%%%%%%%%%%%%%%%%%%%%%%%%%%%%%%%%%%%%%%%%
\section{Conclusions}\label{Sec7}
We investigate the acceleration of the Universe's expansion within the framework of Horava-Lifshitz Gravity, focusing on Viscous Modified Chaplygin Gas (VMMG) and Generalized Cosmic Chaplygin Gas (GCCG) models. We constrain all cosmological parameters using late-time datasets, including BAO, Hubble Measurements obtained from CC Method, SNIa, Q, GRBs, and the recent measurement of the Hubble constant. We select 17 BAO points from an extensive dataset comprising 333 BAO data points, deliberately chosen to minimize correlations. To reduce errors in the posterior distribution caused by correlations between measurements, we use a method to simulate random correlations in the covariance matrix. After verifying the results, we find that these introduced correlations do not significantly change the cosmological parameters. We extract the present-day Hubble function ($H_{0}$) with sound horizon ($r_d$) in both cosmological models. The VMMG model yields estimates of $H_0 = 66.397928 \pm 1.628310 \ \mathrm{km/s/Mpc}$ and $r_d = 145.962162 \pm 2.284883 \ \mathrm{Mpc}$. In contrast, the GCCG model provides slightly different values: $H_0 = 69.234519 \pm 1.043849 \ \mathrm{km/s/Mpc}$ and $r_d = 147.187439 \pm 1.806239 \ \mathrm{Mpc}$. These results indicate a convergence with early Planck estimates for $H_0$ and $r_d$ based on low-redshift measurements. A comparative study across $\Lambda$CDM, VMMG, and GCCG models reveals their distinct evolution patterns of cosmographic parameters. Each model exhibits transitions from deceleration to acceleration, maintaining consistency with theoretical predictions. These frameworks offer varied insights into cosmic dynamics, with numerical values summarized in Table \ref{tab_3}. The diagnostic parameters reveal intriguing dynamics in both VMMG and GCCG models. VMMG transitions from quintessence to Chaplygin gas dominance, while GCCG exhibits similar behavior. High redshifts show phantom behavior in both models, shifting to quintessence at intermediate redshifts. The intricate interplay between these regions underscores the complexity of dark energy models. While the $\Lambda$CDM model exhibits the best fit based on AIC, both VMMG and GCCG models remain viable according to AIC and BIC. Further, the reduced chi-square statistics show all models closely align with observed data, indicating satisfactory goodness of fit. In conclusion, the investigation into the VMMG and GCCG models offers valuable insights that enhance and challenge the conventional $\Lambda$CDM model. These models provide nuanced perspectives on cosmic expansion and evolution, particularly in terms of transitioning from deceleration to acceleration. They yield estimates for the Hubble constant and sound horizon that converge with early Planck results but also reveal distinctive evolution patterns of cosmographic parameters. While the $\Lambda$CDM model currently provides the best fit, the VMMG and GCCG models present viable alternatives, maintaining consistency with observed data and offering new avenues for exploring dark energy dynamics.

\section*{Acknowledgments}
   The author S. K. Maurya appreciates the administration of the University of Nizwa in the Sultanate of Oman for their unwavering support and encouragement.

\bibliographystyle{ieeetr}
\bibliography{mybib}

\end{document}